\documentclass[12pt]{article}
\usepackage{amsmath,amssymb,bbm}
\usepackage{amsmath}
\begin{document}

\begin{center}

{\bf \Large {Remark on Calabi-Yau vacua of the string theory\\
\vskip 0.2cm and the cosmological constant problem}}

\end{center}

\vskip 0.5cm
\begin{center}
{\bf Eun Kyung Park\footnote{E-mail: ekpark1@dau.ac.kr}
and Pyung Seong Kwon\footnote{E-mail: bskwon@ks.ac.kr}\\
\vskip 0.2cm}
\end{center}

\begin{center}
{$^1$Department of Materials Physics, Dong-A University,\\
Busan 604-714, Korea \\

\vskip 0.1cm

$^2$Department of Physics, Kyungsung University,\\
Busan 608-736, Korea}
\end{center}

\thispagestyle{empty}

\vskip 0.2cm
\begin{center}
{\bf Abstract}
\vskip 0.1cm
\end{center}

In the first part of the paper we study solitonic properties of the Calabi-Yau vacua of the string theory. We observe that the Calabi-Yau threefolds of the string theory may be thought of as NS-NS objects whose masses are proportional to $1/g_s^{2}$. In the second part, which is the main part of this paper, we propose, based on the viewpoint that our three-dimensional space is a stack of BPS D3-branes located at the conifold singularity of the Calabi-Yau threefold, a new mechanism to address the cosmological constant problem in the framework of the conventional compactifications where the $n$-form fluxes including NS-NS three-form are all turned off. In this mechanism the four-dimensional cosmological constant $\lambda$ appears as two types, NS-NS type and R-R type, of vacuum energies on the brane plus supersymmetry breaking term, which constitute a brane action density ${\hat I}_{\rm brane}$, and sum of these three terms of ${\hat I}_{\rm brane}$ are forced to vanish by field equations so that $\lambda$ adjusts itself to zero as a result. Also in this mechanism the $d=4$ supersymmetry is broken in the brane region, while still maintaining $\lambda =0$. The supersymmetry breaking occurs as a result of the gauge symmetry breaking of the R-R four-form arising at the quantum level. The substance of the supersymmetry breaking term is a vacuum energy density (of the brane region) arising from the quantum excitations with components along the transverse directions to the D3-brane. We generalize the above mechanism to the case of the flux compactifications where the fluxes are all turned on to stabilize the moduli. In the generalized theory $\lambda$ appears as ${\hat I}_{\rm brane}$ plus the scalar potential ${\mathcal V}_{\rm scalar}$ for the moduli, in contrast to the case of the ordinary flux compactifications where $\lambda$ is simply given by ${\mathcal V}_{\rm scalar}$. Also in this theory any nonzero ${\mathcal V}_{\rm scalar}$ arising from perturbative or nonperturbative corrections is gauged away by the gauge arbitrariness of ${\hat I}_{\rm brane}$ and the condition $\lambda=0$. So $\lambda$ is again expressed only as a brane action density as before, or it simply vanishes by the cancelation between ${\hat I}_{\rm brane}$ and ${\mathcal V}_{\rm scalar}$.

\vskip 0.25cm
\begin{center}
\end{center}

\newpage
\setcounter{page}{1}
\setcounter{footnote}{0}

\baselineskip 6.0mm

\vskip 1cm
\hspace{-0.65cm}{\bf \Large I. Introduction}
\vskip 0.5cm

Background vacua of the ten-dimensional superstring theory are described by a product of the maximally symmetric four-dimensional spacetime and the internal dimensions compactified on a Calabi-Yau threefold. Such a compactification is appropriate because it admits $SU(3)$ holonomy which yields unbroken ${\mathcal N}=1$ supersymmetry in four dimensions. The moduli spaces of Calabi-Yau manifolds usually contain certain limit points, the conifolds, which are identified as transition points where the moduli spaces of two distinct Calabi-Yau manifolds meet \cite{1, 1-1}. Thus the conifolds commonly occur in the moduli spaces of Calabi-Yau manifolds in such a way that each moduli space of the Calabi-Yau manifold generally contains a single point corresponding to the conifold, and the geometry of this conifold is relatively simple to the other Calabi-Yau spaces.\footnote{Such an each moduli space of the Calabi-Yau manifold is usually taken as a single Calabi-Yau space.} This is why string-inspired brane world models have considered configurations of D3-branes located at conifold singularities \cite{2,3}.

In brane world models a stack of D3-branes is identified with our three-dimensional external space, which is assumed to be dynamical. Recently, however, there was a conjecture that fundamental background brane immanent in our spacetime may perhaps be NS-NS type brane, rather than D-brane. In \cite{4} it was argued that in ($p+3$)-dimensional string theory the existence of NS-NS type $p$-brane is essential to obtain background geometries $R_2$ or $R_2/{\mathbb Z}_n $ on the transverse dimensions, and the usual codimension-2 brane solutions with these background geometries already contain NS-NS type brane implicitly in their ansatz. Similar thing happens in the case of codimension-1 brane solutions as well. In \cite{5} the authors have studied codimension-1 brane solutions of the five-dimensional models compactified on $S_1 /{\mathbb Z}_2$. They showed that in string theoretical setup the existence of the background NS-branes are indispensable to obtain flat geometry $M_4 \times S_1 / {\mathbb Z}_2$, and without these branes the five-dimensional metric becomes singular everywhere.

In these lines of study it would be important to check the case of the ten-dimensional full-fledged string theory\footnote{Aside from the F-theory this may correspond to the theory of codimension-$n$ branes with $n \geq 3$.} as a final confirmation of the given conjecture. Indeed Calabi-Yau vacua of the ten-dimensional string theory give an indication of the similar behavior associated with NS5-branes. Remarkably, there were observations \cite{3,6} that two intersecting NS5-branes can be thought of as a T-dual configuration of the conifold singularity. For instance in \cite{3} the authors have considered two IIA configurations of D4-brane suspended between two NS5-branes, i.e., one with two parallel NS5-branes and another with two orthogonal NS5-branes. They have qualitatively shown that under T-duality the former corresponds to a IIB metric of D3-brane plus a Taub-NUT space in the transverse directions, while the later corresponds to that of D3-brane at a conifold singularity. In the absence of D3-brane the latter case implies that the conifold singularity is T-dual to a configuration of two orthogonal NS5-branes.

A pair of T-dual configurations are just two different geometrical realizations (in the Calabi-Yau target space) of the same conformal field theory, and in this sense they are recognized to be physically equivalent even though they are topologically distinct in general. Thus in the framework of the conformal field theory the conifold singularity of the Calabi-Yau space becomes equivalent to the background configuration with two intersecting NS5-branes, and since the moduli spaces of the Calabi-Yau manifolds always contain conifold singularities we are inclined to say that the NS5-branes are involved at any rate in the background vacua of the string theory and further the background vacua of the string theory may include NS5-branes implicitly in their Calabi-Yau ansatz. Indeed in Sec. III we observe that the generic compact Calabi-Yau threefolds of the string theory contain $n$ couples of intersecting Kaluza-Klein (KK) monopoles, the T-dual counterparts of the NS5-branes, at the singularities and each of these intersecting KK monopoles can be thought of as an NS-NS type soliton with mass proportional to $1/g_s^2$.

The first part of this paper is mainly concerned with this issue though it is continued to the cosmological constant problem in the second part. In the second part, which is the main part of this paper, we propose a new mechanism to address the cosmological constant problem, based on the viewpoint that our three-dimensional space is a stack of BPS D3-branes located at the conifold singularity of the Calabi-Yau threefold. In Secs. IV-VII, we first consider the case of the conventional compactifications where the $n$-form fluxes including NS-NS three-form are all turned off. In this case we find that the four-dimensional cosmological constant $\lambda$ appears as two types, NS-NS type and R-R type, of vacuum energies on the brane plus the supersymmetry breaking term, which constitute a brane action density ${\hat I}_{\rm brane}$, and sum of these three terms of ${\hat I}_{\rm brane}$ are forced to vanish by field equations so that $\lambda$ adjusts itself to zero as a result. Also in this mechanism the $d=4$ supersymmetry is broken in the brane region, while still maintaining $\lambda=0$. The supersymmetry breaking occurs as a result of the gauge symmetry breaking of the R-R four-form arising at the quantum level. So the brane region is locally anomalous. But the total anomaly of the brane region turns out to vanish by the condition $\lambda=0$.

The primary cause of the supersymmetry breaking is these anomalies for the string fields with support on the D3-brane, and the substance of the supersymmetry breaking term in the action is a vacuum energy density (of the brane region) arising from the quantum excitations with components along the transverse directions to the D3-brane. In Sec. IX it is argued that the supersymmetry breaking in the conventional compactifications gives a mass to the dilaton which is estimated to be $m_{\Phi}^2 \sim g_s m_s^2$, where $m_s$ is the fundamental mass scale of the string theory. Also since $m_{\Phi}$ can be roughly identified with $m_{\rm sp}$, the typical mass scale of the Standard Model superpartners, one obtains $m_{\rm sp}^2 \sim g_s m_s^2$ from the above equation.

It is also argued in Sec. 7.2 that the configuration with broken supersymmetry is more favored than the other with an unbroken supersymmetry. These two configurations are equally qualified for a solution to the field equations, but the former is more favored by the action principle than the latter because the former takes lower values of the total action than the latter. For this matter a different possible viewpoint is also briefly presented at the end (the fourth last paragraph) of Sec. IX.

In Sec. VIII we finally generalize the above mechanism to the case of the flux compactifications where the fluxes are all turned on to stabilize the moduli. In the generalized theory $\lambda$ appears as a sum of two terms, ${\hat I}_{\rm brane}$ and the scalar potential ${\mathcal V}_{\rm scalar}$ for the moduli, in contrast to the case of the ordinary flux compactifications where $\lambda$ is simply given by ${\mathcal V}_{\rm scalar}$. Among these two terms ${\hat I}_{\rm brane}$ depends on gauge parameters and therefore it is arbitrary. Beside this, it is shown in Sec. 8.3 that $\lambda$ is always required to vanish by field equations. So from all this one finds that any nonzero ${\mathcal V}_{\rm scalar}$ arising from the perturbative or nonperturbative corrections is gauged away by the gauge arbitrariness of ${\hat I}_{\rm brane}$ and the condition $\lambda =0$. As a result $\lambda$ is again expressed only as a brane action density, or it simply vanishes by the cancelation between ${\mathcal V}_{\rm scalar}$ and ${\hat I}_{\rm brane}$.

\vskip 1cm
\hspace{-0.65cm}{\bf \Large II. Conifold as an NS-NS soliton}
\vskip 0.5cm
\setcounter{equation}{0}
\renewcommand{\theequation}{2.\arabic{equation}}

Consider a configuration of background fields $G_{MN}$, $\Phi$ and $B_{MN}$ of the NS-NS sector. The target space action for these background fields is given by
\begin{equation}
I_{10} = \frac{1}{2 \kappa_{10}^2} \int d^{10} x \sqrt{-G}\,\, e^{-2 \Phi} \Big[{\mathcal R}_{10} + 4(\nabla \Phi)^2 - \frac{1}{2 \cdot 3 !}\, H_3^2 \Big]\,\,,
\end{equation}
where $H_3$ ($\equiv dB_2$) is the field strength of the NS-NS two-form $B_{MN}$. In the absence of NS5-branes $B_{MN}$ and consequently $H_3$ all vanish.  In this configuration $I_{10}$ admits Ricci-flat solutions and one of which takes the form $ds_{10}^2 = ds_{0123}^2 + ds_{\rm conifold}^2$ where $ds_{\rm conifold}^2$ represents the conifold metric
\begin{equation}
ds_{\rm conifold}^2 = dr^2 + r^2 d \Sigma_{1,1}^2 \,\,,
\end{equation}
where
\begin{equation}
d \Sigma_{1,1}^2 = \frac{1}{9} (d \psi +  \sum_{i=1}^{2} \cos \theta_i d \phi_i )^2 + \sum_{i=1}^{2} \frac{1}{6} (d \theta_i^2 +  \sin^2 \theta_i d \phi_i^2 ) \,\,
\end{equation}
is an Einstein metric representing the base of the cone.

Under T-duality along the isometry direction $\psi$, (2.2) turns into
\begin{equation}
ds_{\rm T-dual}^2 = dr^2 + \frac{9}{r^2} d \psi^2 + r^2 \sum_{i=1}^{2} \frac{1}{6} (d \theta_i^2 +  \sin^2 \theta_i d \phi_i^2 ) \,\,
\end{equation}
plus two-form field contribution $B_{\psi \phi_i}$ which is given by $\cos \theta_i$. The T-dual metric (2.4) cannot be Ricci-flat because the right-hand side of the Einstein equation contains a matter field contribution arising from $B_{\psi \phi_i}$. (2.4) is not only non-Ricci-flat, but it is very singular. The scalar curvature calculated from (2.4) is given by ${\mathcal R}=16/r^2$, which goes to infinity as $r$ goes to zero. Thus the T-duality transformation shows that the conifold metric is equivalent to a sum of non-Ricci-flat singular metric and NS5-branes described by $B_{\psi \phi_i}$, which is reminiscent of the codimension-1 \cite{5} and codimension-2 \cite{4} brane world models where each of the flat geometries of the transverse spaces can be formally expressed as a sum of singular metric and NS-NS type brane.

Though (2.4) is T-dual to (2.2) it breaks the supersymmetry completely. In general the localized metric of the form (2.2) does not preserve the supersymmetry under T-duality transformation \cite{7}. In many cases, and in particular if we want supersymmetric solution, it is more convenient to consider the T-dual configurations of smeared NS5-branes. Let us consider a configuration of intersecting $n$ NS5-NS5$^{\prime}$-branes extended along NS5=(012345) and NS5$^{\prime}=(012389)$ respectively and smeared except for one overall transverse direction, $x^7$.
This configuration preserves $1/4$ supersymmetries \cite{8} and the metric takes the form:
\begin{equation}
ds_{\rm NS-NS^{\prime}}^2 = ds_{0123}^2 + H_{\rm NS}^2 ds_{67}^2 + H_{\rm NS} (ds_{45}^2 + ds_{89}^2 ) \,\,,
\end{equation}
where $H_{\rm NS}$ is the harmonic function for the $n$-coincident NS5-branes, $H_{\rm NS} = 1+ {n}|x^{7}|$, and the NS-NS three-form field strengths are given by $H_{645}=H_{689}=n$. Under T-duality along $x^6$ it turns into delocalized metric of the conifold. Omitting $ds_{0123}^2$ it reads
\begin{equation}
ds_{\rm conifold}^2 = H_{\rm NS}^2 ds_7^2 + H_{\rm NS}^{-2} (ds_{6} + B_{64} ds_4 +B_{68} ds_8 )^2 + H_{\rm NS} (ds_{45}^2 + ds_{89}^2 ) \,\,,
\end{equation}
where $B_{46}=nx^5$ and $B_{86}=nx^9$.

The metric (2.6) suggests that the conifold geometry is due to an NS-NS type extended object because it contains the harmonic function for the NS5-branes which originally appears in the metric for the NS5-brane. To see this more precisely consider a simpler configuration of a single NS5-brane extended along ($012345$)-directions:
\begin{equation}
ds_{\rm NS}^2 = ds_{012345}^2 + H_{\rm NS} ds_{6789}^2 \,\,
\end{equation}
and nonzero antisymmetric background $B_{6i}$ ($i=7,8,9$), where $H_{\rm NS}= 1+1/r^2$. Now we compactify $x^6$ to take T-duality. Under T-duality along $x^6$, (2.7) turns into \cite{9}
\begin{equation}
ds_{\rm KK}^2 = ds_{012345}^2 + ds_{\rm Taub-NUT}^2 \,\,,
\end{equation}
where $ds_{\rm Taub-NUT}^2$ is the four-dimensional Taub-NUT metric
\begin{equation}
ds_{\rm Taub-NUT}^2 = H_{\rm NS}^{-1} ( ds_{6} + \omega_i ds_i )^2 + H_{\rm NS} ds_{789}^2 \,\,,
\end{equation}
where $\omega_i =B_{6i}$ with $\vec\nabla \times \vec\omega = \pm \vec \nabla H_{\rm NS}$, but $H_{\rm NS}$ is now given by $H_{\rm NS}= 1+1/r$ because the number of noncompact transverse directions has been reduced to three in (2.9). (2.8) describes the Kaluza-Klein monopole which can be identified as a five-dimensional object extended along ($012345$). Also since it contains $H_{\rm NS}$, we suspect that the KK monopole is also an NS-NS type five-brane just like the NS5-brane.

To convince ourselves that the KK monopole is really an NS-NS object let us calculate the mass of the KK monopole described by (2.8). For the metric (2.8), the first term in (2.1) can be converted into an action for the five-dimensional KK monopole:
\begin{equation}
I_{5{\rm KK}} = \frac{1}{16 \pi G_{\rm K}} \int d^5 x \sqrt{-g_5} \, e^{-2 \hat{\Phi}}\, {\mathcal R}_5 \,\,,
\end{equation}
where ${16 \pi G_{\rm K}} = {2 \kappa_{10}^2} \, {g_s^2}/{\rm{Vol}({\rm NS}5)}$ with $\rm{Vol}({\rm NS}5)$ being volume of the five-brane extended along ($12345$), and $\hat{\Phi}$ is defined by $e^{\hat{\Phi}}=e^{\Phi}/g_s$ so that $e^{\hat{\Phi}} \rightarrow 1$ as $r \rightarrow \infty$. (2.10) is identical with the action in \cite{10} except that $\sqrt{-g_5}$ is replaced by $\sqrt{-g_5}\, e^{-2 \hat{\Phi}}$. The mass of the monopole per unit volume of the five-brane is therefore
\begin{equation}
m = \frac{1}{(2\pi)^5} \, \frac{m_s^8 R^2}{g_s^2} \,\,,
\end{equation}
where $m_s$ is the string mass scale defined by $2 \kappa_{10}^2 = (2 \pi)^7 / m_s^8$, and $R$ is the radius of the compact dimension $ds_6$. (2.11) shows that the mass density $m$ of the monopole is proportional to $1/g_s^2$ as expected, which suggests that the KK monopole is an NS-NS type soliton just like the NS5-brane.

Turning back to (2.8), introduce a D3-brane with world-volume along ($0123$) on the geometrical singularity of the Taub-NUT space. The metric for this configuration will be (see \cite{3}):
\begin{equation}
{d}s_{\rm KK-D}^2 = H_{\rm D}^{-1/2} ds_{0123}^2 +  H_{\rm D}^{1/2} [ ds_{45}^2 + H_{\rm NS}^{-1} (ds_{6} + \omega_i ds_i )^2 + H_{\rm NS} ds_{789}^2 ] \,\,,
\end{equation}
where $H_{\rm D}$ is the harmonic function for the D3-brane. The harmonic function for the D$p$-brane, $H_{\rm D} = 1+ g_s /r^{7-p}$, contains an additional parameter $g_s$ as compared with that for the NS5-brane.\footnote{More generally the harmonic function for the $n$ coincident D$p$-branes takes the form $H_{\rm D} = 1+ n g_s /r^{7-p} +$ the terms linear in $g_s$, where we have set $\alpha^{\prime} =1$} Hence in the limit $g_s \rightarrow 0$, the effect of D-brane on the background geometry disappears and the geometry of the transverse dimensions is entirely determined by the configuration of NS5-branes. Indeed (2.12) reduces to (2.8) in the limit $g_s \rightarrow 0$ which again shows that the (extended source of the) KK monopole is a pure NS-NS type 5-brane whose mass density is proportional to $g_s^{-2}$ just like the NS5-brane. The above example shows that changes of internal geometries caused by D-branes necessarily contain the factor $g_s$ in the metric through the harmonic function $H_{\rm D}$, and conversely if the geometry does not contain $g_s$ in the metric we may suspect that it could be caused by NS-NS type objects.

The same story goes on for the conifold case too. The configuration of a D3-brane at the conifold singularity is given by
\begin{equation}
ds_{\rm conifold-D}^2 = H_{\rm D}^{-1/2} ds_{0123}^2 + H_{\rm D}^{1/2} [H_{\rm NS}^2 ds_{7}^2 + H_{\rm NS}^{-2} (ds_{6} +B_{64} ds_4 + B_{68} ds_8 )^2 + H_{\rm NS} (ds_{45}^2 + ds_{89}^2 )]\,\,.
\end{equation}
In the limit $g_s \rightarrow 0$ (2.13) reduces to (2.6). So we suspect that the conifold geometry described by (2.6) may be thought of as being caused by NS-NS objects because it does not contain $g_s$ in the metric. The metric (2.6) can be converted back into the localized metric by replacing the Cartan basis 1-forms of $R_2 \times R_2$ by those of $S_2 \times S_2$ (see \cite{8}):
\begin{equation}
dx^{4,8} \rightarrow \sin \theta_{1,2} d \phi_{1,2}  \,\,, ~~~~~~~~~dx^{5,9} \rightarrow d\theta_{1,2}   \,\,.
\end{equation}
As a result (2.6) becomes (2.2) which is also independent of $g_s$, suggesting that the conifold geometry (2.2) is also caused by NS-NS objects.

\vskip 1cm
\hspace{-0.65cm}{\bf \Large {III. Calabi-Yau threefolds as NS-NS objects}}
\vskip 0.5cm
\setcounter{equation}{0}
\renewcommand{\theequation}{3.\arabic{equation}}

Since the conifold is T-dual to intersecting NS5-branes and KK monopole is T-dual counterpart of NS5-brane, one naturally expects that the conifold may be identified with a configuration of intersecting KK monopoles. This is indeed the case. The authors of \cite{3} observed that the conifold can be thought of as an ALE fibration over a ${\mathbb P}_1$, where the fibers are given by a family of ALE spaces parameterized by the coordinate of ${\mathbb P}_1$. In complex coordinates the conifold is described by a quadric in ${\mathbb C}_4$:
\begin{equation}
({z_1})^2 + ({z_2})^2 + ({z_3})^2 + ({z_4})^2 = 0 \,\,.
\end{equation}
This equation can be rewritten as
\begin{equation}
\prod_{i=1}^{2} (z_1 - \zeta_i )^2+ z_2^2 + z_3^2 =0
\end{equation}
with $\zeta_i$ given by $\zeta_1 = \zeta$ and $\zeta_2 =-\zeta$, where $\zeta=iz_4$. (3.2) describes an $A_1$ ALE space $R_4 /{\mathbb Z}_2$ (blown up by ${\mathbb P}_1$) which is just the Eguchi-Hanson (EH) space represented by the metric \cite{11}
\begin{equation}
ds_{\rm EH}^2 = (1- \frac{a^4}{r^4})^{-1} dr^2 + \frac{r^2}{4} (1- \frac{a^4}{r^4})(d\psi +\cos\theta d \phi)^2 + \frac{r^2}{4} (d \theta^2 + \sin^2 \theta d \phi^2 ) \,\,,
\end{equation}
where the size $a^2$ of the EH space depends linearly on $\zeta(=iz_{4})$.

In a new coordinate system (3.3) can be transformed into a two center Gibbons-Hawking metric \cite{12}, which however is identical with two-center Taub-NUT space near the singularity of the ALE space. Also the distance between two centers in the Taub-NUT space varies linearly as a function of $z_4$ since it is linearly proportional to $a^2$ (see \cite{12}), and therefore it varies linearly as a function of $z_4$ and the locus of these two centers, which is a set of two sections on the fibered space $R_4/{\mathbb Z}_2 \times {\mathbb P}_1$, may be identified as two intersecting KK monopoles. In this way one finds that near the singularity the conifold can be identified with two intersecting KK monopoles which intersect at the conifold singularity given by $z_1 =z_2 = z_3 =z_4 =0$. This confirms the conjecture that the conifold geometry is caused by an NS-NS type extended source.

Now let us turn to the Calabi-Yau threefolds. As mentioned already the moduli spaces of Calabi-Yau manifolds usually contain conifolds as transition points where moduli spaces of two distinct Calabi-Yau manifolds meet. Indeed the conifold itself is a singular noncompact Calabi-Yau threefold, and we might say that a set of $N$ such conifolds join with some other regions to constitute a regular compact Calabi-Yau manifold in the sense that the generic Calabi-Yau threefolds usually contain singularities which locally look like the conifold. In order to see this a little more in detail let us briefly review some of the first section of \cite{1}.

The simplest example of the nonsingular CICY threefold is ${{\Bigg\lbrack \begin{matrix} 4\,\,\vline\,\vline\,\, 4~1 \\ 1\,\,\vline\,\vline\,\, 1~1 \end{matrix} \Bigg \rbrack}}_{-168}^2$. This is a projective variety defined by the homogeneous equations
\begin{equation}
X(x) y_1 + U(x) y_2 =0\,\,, ~~~~~~~~~~V(x) y_1 + Y(x) y_2 =0\,\,,
\end{equation}
where $X$, $U$ and $V$, $Y$ are any general quartic and linear polynomials in the homogeneous coordinates of ${\mathbb P}_4$ respectively. ${{\Bigg\lbrack \begin{matrix} 4\,\,\vline\,\vline\,\, 4~1 \\ 1\,\,\vline\,\vline\,\, 1~1 \end{matrix} \Bigg \rbrack}}_{-168}^2$ is nonsingular because it is a small resolution of ${\Big \lbrack 4\,\,\vline\,\vline\,\, 5 \Big \rbrack}_{-200}^1$ which is defined by a quintic hypersurface in ${\mathbb P}_4$. Indeed for nonvanishing homogeneous coordinates, $y$'s, of ${\mathbb P}_1$, (3.4) implies
\begin{equation}
XY-UV=0 \,\,,
\end{equation}
which defines the simplest Calabi-Yau threefold ${\Big \lbrack 4\,\,\vline\,\vline\,\, 5 \Big \rbrack}_{-200}^1$. ${\Big \lbrack 4\,\,\vline\,\vline\,\, 5 \Big \rbrack}_{-200}^1$ is singular and has a number of isolated nodes at the points where $X(x)$, $Y(x)$, $U(x)$, $V(x)$ all vanish simultaneously. For generic $X$, $Y$, $U$, $V$ the number of nodes is 16 because $X$ and $U$ are quartic polynomials.

The above $X$, $Y$, $U$, $V$ can be taken as coordinates in ${\mathbb C}_4$ describing the conifold with singularity being located at $X=Y=U=V=0$. Thus at each of $16$ nodes the conifold singularity develops, and the neighborhoods of these 16 points can be identified as conifolds which smoothly join to the main body to constitute the compact Calabi-Yau ${\Big \lbrack 4\,\,\vline\,\vline\,\, 5 \Big \rbrack}_{-200}^1$. Indeed the generic compact Calabi-Yau threefolds usually contain certain numbers of conifolds at the singularities in such a way that the entire topology is characterized by the Hodge number $h^{1,1}$ and some negative values of $\chi$. Also since each conifold is made of two intersecting KK monopoles we finally observe that the Calabi-Yau threefolds are also NS-NS objects in the sense that they contain numbers of NS-NS solitons with mass proportional to $1/g_s^2$.

\vskip 1cm
\hspace{-0.65cm}{\bf \Large {IV. The four-dimensional cosmological constant}}
\vskip 0.5cm
\setcounter{equation}{0}
\renewcommand{\theequation}{4.\arabic{equation}}

The fact that the conifolds of the Calabi-Yau threefolds of the string theory can be thought of as NS-NS solitons with mass proportional to $1/g_s^2$ is important in addressing the cosmological constant problem. The cosmological constant problem is one of the most mysterious problems in the area of the theoretical high energy physics. Though it is very complicated problem \cite{13}, its main point may be simply summarized as why the cosmological constant $\lambda$ of the four-dimensional spacetime is so small despite the enormous contributions to the vacuum energy arising from the quantum fluctuations of SM fields.

One of the most promising candidates for the solution to this problem may be found from the brane world models where the intrinsic curvature of the brane is a priori zero. But in these models the whole vacuum energy including quantum fluctuations of SM-fields always contributes to changing the internal geometry because the geometry of the four-dimensional spacetime is already fixed to have a zero cosmological constant from the beginning \cite{14,15}. Thus the geometry of the internal space is now expected to be severely disturbed by the quantum fluctuations, which then leads to a severe disturbance of the observed coupling constants. This problem may be naturally solved by accepting the viewpoint that the conifolds of the internal Calabi-Yau space are NS-NS solitons made of two intersecting KK monopoles which themselves are NS-NS solitons. As is the case with the usual solitons the NS-NS solitons are very heavy and rigid because their masses are proportional to $1/g_s^2$, and as a result the effect of the vacuum fluctuations exerting on the internal geometry is highly suppressed by the factor of $g_s^2$ in the equations of motion \cite{16}.

Though the solitonic interpretation of the Calabi-Yau space has an important consequence, it alone is not enough to explain the smallness of the cosmological constant completely. Because the four-dimensional Planck mass, $M_{pl}$, is inversely proportional to $g_s$ (see (4.27)), the $g_s^2$-suppression caused by the heaviness of the NS-NS type soliton only suggests a value $\sim (\mu^4 /M_{pl}^2 )$ for $\lambda$, where $\mu$ is a mass scale at which the symmetry of the theory is broken. So if we take the supersymmetry on the brane as a broken symmetry of the theory, $\mu$ will be about\footnote{But see also the Sec. IX where $\mu$ ($> m_{\rm sp}$) is estimated to be $\mu^2 > g_s m_s^2$, according to which $\mu$ could be much larger than the conventional scale of order $\sim TeV$.} $\sim TeV$ and therefore ($\mu^4 /M_{pl}^2$) takes the value $\sim (TeV)^4 /(10^{19} GeV)^2$, which is certainly too large to be a correct value for the present $\lambda$. In order to obtain much smaller value for $\lambda$ we may need an entirely new additional mechanism. In the following sections we propose a new mechanism with which to solve the cosmological constant problem. As the beginning of the discussion we first consider the conventional compactifications where the $n$-form fluxes stabilizing the moduli are all turned off.

Consider a configuration of D3-brane extended along ($0123$) at the conifold singularity,\footnote{In the following discussion, for simplicity, we will consider a configuration with a single D3-brane, instead of a stack of $n$-coincident D3-branes, located at the conifold singularity. But there is no essential difference between these two cases and the extension is trivial.} and assume that the D3-brane is basically a BPS state. The total action for this configuration is given by the sum $I_{\rm total} = I_{\rm bulk} + I_{\rm brane}$ with
\begin{equation}
I_{\rm bulk} = \frac{1}{2\kappa_{10}^2} \int d^{10}x \sqrt{-G} \, \Big[ e^{-2\Phi} [{\mathcal R}_{10} + 4 (\nabla \Phi)^2 ] - \frac{1}{2 \cdot 5!}\, {\mathcal F}_5^2 \Big] \,\,,
\end{equation}
and
\begin{equation}
I_{\rm brane} = - \int d^4 x \sqrt{-\det|G_{\mu\nu}|} \,T(\Phi) + \mu_0 \int  A_4 \,\,,
\end{equation}
where ${\mathcal F}_5$ is a self-dual five-form, the field strength of the R-R four-form $A_4$, and $G_{\mu \nu}$ is a pullback of $G_{MN}$ to the four-dimensional brane world. Also $T(\Phi)$ represents the tension of the D3-brane, so at the tree level it is simply $T(\Phi)=T_0 \, e^{-\Phi}$. But if we include quantum corrections it becomes $T(\Phi)=T_0 \, e^{-\Phi} + \rho_{\rm vac} (\Phi)$, where $\rho_{\rm vac} (\Phi)$ represents the quantum correction terms and it is identified with the (NS-NS sector) vacuum energy density of the three-dimensional space. Similarly, $\mu_0$ represents the R-R charge of the D3-brane, which is electric under $A_4$. If we include quantum corrections $\mu_0$ turns into $\mu (\Phi)$ which is $\mu (\Phi)=\mu_0 + \delta \mu (\Phi)$ where $\delta \mu (\Phi)$ is an R-R counterpart of $\rho_{\rm vac}(\Phi)$ and therefore represents the quantum correction terms. See case III of Sec. V for the details.

Now we introduce a general ansatz for the ten-dimensional metric as
\begin{equation}
ds_{10}^2 = e^{A({\hat r})} ds_6^2 + e^{B({\hat r})} g_{\mu\nu}(x) dx^{\mu} dx^{\nu} \,\,,
\,\,
\end{equation}
where
\begin{equation}
ds_6^2= d{\hat r}^2 + R^2 ({\hat r})\,d\Sigma_{1,1}^2 \equiv h_{m n}(y) dy^m dy^n
\end{equation}
is the metric of the internal dimensions, while $g_{\mu\nu}(x) dx^{\mu} dx^{\nu}$ is the metric of the four-dimensional spacetime. In the above metric $e^{A({\hat r})}$ is an extra degree of freedom which could have been absorbed into $ds_6^2$, so it can be taken arbitrarily as we wish. The ansatz for the R-R four-form, on the other hand, is given by
\begin{equation}
A_4 = \xi ({\hat r}) \sqrt{- g_4} \,\, dt \wedge dx^1 \wedge dx^2 \wedge dx^3
\end{equation}
where $g_4$ is the determinant of the four-dimensional metric $g_{\mu\nu}(x)$. (4.5) is an appropriate ansatz for $A_4$ because it is consistent with the homogeneous and isotropic geometry of the four-dimensional spacetime.

As mentioned above, ${\mathcal F}_5$ is a (anti-) self-dual five-form, and we may write it as
\begin{equation}
{\mathcal F}_5 = F_5 \pm i\, {}^{\ast} F_5 \,\,,
\end{equation}
where $F_5$ is the field strength of $A_4$:
\begin{equation}
F_5 =dA_4 = \sqrt{-g_4}\, (\partial_{\hat r} \xi)\, dr \wedge dt \wedge dx^1 \wedge dx^2 \wedge dx^3 \,\,,
\end{equation}
and ${}^{\ast} F_5$ is its dual:
\begin{equation}
{}^{\ast}F_5 = - e^{2A-2B} \sqrt{h_6} \,(\partial_{\hat r} \xi)\, d\psi \wedge d\theta_1 \wedge d\phi_1 \wedge d\theta_2 \wedge d\phi_2 \,\,,
\end{equation}
where $h_6$ is the determinant of $h_{mn}(y)$. We see that (4.6) satisfies the (anti-) self-duality condition $i\, {}^{\ast} {\mathcal F}_5 = \pm {\mathcal F}_5$, but at this point we are not allowed to do this. Imposing self-duality condition on the action would result in wrong field equations. When ${\mathcal F}_5$ is given by (4.6), ${\mathcal F}_5^2$ contains two terms, one from $F_5$ and another from ${}^{\ast}F_5$. These two terms take the same form when they are expressed in terms of $\partial_{\hat r} \xi$, and this gives twice the term $F_5^2$ in the action, which will be a double consideration and will lead to wrong field equations. A way out of this difficulty is to set ${\mathcal F}_5^2$ simply as ${\mathcal F}_5^2 = F_5^2$ in the action (4.1) and postpone imposing the self-duality condition until we find the whole solutions to the field equations \cite{17}. Thus we impose the self-duality condition on the solution (not on the action) simply as a supplementary constraint, and in this case the dynamics of ${}^{\ast}F_5$ is governed by the dual action which is given by (4.1) plus (4.2) but where $F_5$ and $A_4$ are replaced by their duals \cite{18}.

Now we set $e^{\Phi}=g_s e^{\hat{\Phi}}$ and choose
\begin{equation}
A\,=\,\hat{\Phi} -B  \,\,.
\end{equation}
Then the total action is converted into
\begin{eqnarray}
I_{\rm bulk} = \frac{1}{2\kappa_{10}^2 \, g_s^2} \Big[\int d^4 x \sqrt{-g_4}\, {\mathcal R}_4 (g_{\mu\nu})\Big] \Big[\int d^6 y \sqrt{h_6} \,\,e^{\hat{\Phi} - 2B} \Big]  ~~~~~~~~\nonumber \\
+\, \frac{1}{2\kappa_{10}^2 \, g_s^2} \Big[\int d^4 x \sqrt{-g_4}\Big]\,\Big[ \int d^6 y \sqrt{h_6}\,\,\big[{\mathcal R}_6 (h_{mn}) - (\partial \hat{\Phi})^2 \nonumber \\
+\, 2(\partial \hat{\Phi})(\partial B) - 2(\partial B)^2 +\frac{g_s^2}{2} e^{2\hat{\Phi} - 4B} (\partial_{\hat r} \xi)^2 \big] \Big]~~~~~~~~~~~~~\,
\end{eqnarray}
plus
\begin{equation}
I_{\rm brane} = \Big[\int d^4 x \sqrt{-g_4} \Big] \Big[ -\int d^6 y \sqrt{h_6}\,\, e^{2B} T(\Phi) \delta^6 (\vec{\hat r}) + \int d^6 y \sqrt{h_6} \,\,\mu (\Phi)\, \xi ({\hat r}) \delta^6 (\vec{\hat r}) \Big] \,\,,
\end{equation}
where the delta function $\delta^6 (\vec{\hat r})$ is defined by $\int d^6 y \sqrt{h_6} \,\,\delta^6 (\vec{\hat r})=1$. We see that each term of the total action appears as a product of four-dimensional and six-dimensional actions and we can obtain four-dimensional and six-dimensional field equations separately from the total action.

Let us first consider the field equations defined on the six-dimensional internal space. They are obtained from the six-dimensional effective action $I_{\rm total}/[\int d^4 x \sqrt{- g_4}\,]$, where $I_{\rm total}$ is given by (4.10) plus (4.11). We have
\begin{equation}
-\mathcal{H} + \frac{g_s^2}{4}\,e^{2\hat{\Phi} -4B}\,{\xi^{\prime}}^2 -\frac{1}{2}\,e^{\hat{\Phi} -2B} \beta+ 10 \big( \frac{{R^{\prime}}^2}{R^2} - \frac{1}{R^2}\,\big)=0 \,\,,
\end{equation}
\begin{equation}
4\frac{R^{\prime\prime}}{R}+\mathcal{H} - \frac{g_s^2}{4}\,e^{2\hat{\Phi} -4B}\,{\xi^{\prime}}^2 -\frac{1}{2}\,e^{\hat{\Phi} -2B} \beta+ 6 \big( \frac{{R^{\prime}}^2}{R^2} - \frac{1}{R^2}\,\big)=0 \,\,,
\end{equation}
\begin{equation}
\frac{1}{R^5}\,(R^5 \hat{\Phi}^{\prime})^{\prime}=2\kappa_{10}^2 \, {g_s^2}\, \Big[ e^{2B} \, \Big(T(\Phi) + \frac{\partial T(\Phi)}{\partial \Phi} \Big) -\frac{\partial \mu (\Phi)}{\partial \Phi} \xi(\hat{r}) \Big] \,\delta^6 (\vec{{\hat{r}}}\,) \,\,,
\end{equation}
\begin{eqnarray}
\frac{1}{R^5}\,(R^5 B^{\prime})^{\prime} \,-\frac{g_s^2}{2}\,e^{2\hat{\Phi} -4B}\,{\xi^{\prime}}^2 - \frac{1}{2}\,e^{\hat{\Phi} -2B}\beta ~~~~~~~~~~~~~~~~~~~~~~~~\nonumber  \\
~~~~~~~~~~~~~~~~= 2\kappa_{10}^2 \, {g_s^2}\,\Big[ e^{2B} \, \Big(T(\Phi) + \frac{1}{2}\frac{\partial T(\Phi)}{\partial \Phi} \Big) -\frac{1}{2} \frac{\partial \mu (\Phi)}{\partial \Phi} \xi(\hat{r}) \Big] \,\delta^6 (\vec{{\hat{r}}}\,) \,\,,
\end{eqnarray}
where the "prime" denotes the derivative with respect to $\hat{r}$, and $\mathcal{H}$ and $\beta$ are defined, respectively, by $\mathcal{H} \equiv \frac{1}{2} {\hat{\Phi}^\prime}{}^2 - {\hat{\Phi}}^{\prime} B^{\prime} + {B^{\prime}}^2$ and $\beta = \Big[{\int d^4 x \sqrt{-g_4}\,{\mathcal R}_4 }\Big] \big/ \Big[{\int d^4 x \sqrt{-g_4}}\Big]$. Among these equations the first two are the Einstein equations,\footnote{Equations (4.12) and (4.13) are, respectively, the $rr$ and $\theta_i \theta_i$ components of the Einstein equation obtained in the orthonormal frame. But in the orthonormal frame the $\phi_i \phi_i$- and $\psi \psi$- component equations are precisely identical to the $\theta_i \theta_i$-component equation, and consequently we have only two independent Einstein equations, (4.12) and (4.13).} while the last two are linear combinations of the equations for $\hat{\Phi}$ and $B$. The Einstein equations do not contain the delta function terms on the right-hand sides. This must be so because $T(\Phi)$ and $\mu (\Phi)$ do not couple to $h_{mn}$ (see (4.2)) and the determinant $\sqrt{h_6}$ introduced in (4.11) is merely artificial. In any case, apart from the field equation for $\xi({\hat r})$ the above four equations constitute a complete set of linearly independent field equations to solve.

The field equation for $\xi({\hat r})$ is given by
\begin{equation}
\frac{1}{R^5}\,\frac{d}{d{\hat r}} \big[e^{2\hat{\Phi} -4B}\, R^5 \,\frac{d\xi}{d{\hat r}} \big] = 2 \kappa_{10}^2 \, \mu (\Phi) \,\delta^6 (\vec{\hat r}\,) \,\,,
\end{equation}
which, upon integration, gives
\begin{equation}
\partial_{\hat r} \xi = \frac{2 \kappa_{10}^2 \, \mu_{D}}{\rm Vol(B)}\, \frac{e^{-2\hat{\Phi} + 4B }}{R^5}\,\,,~~~~~\Big(\mu_{D} \equiv \mu(\Phi(0)) \Big)\,\,,
\end{equation}
where ${\rm Vol(B)}$ represents the volume of the base of the cone with unit radius: ${\rm Vol(B)}=\int \epsilon_5$ with $\epsilon_5 = \sqrt{\det|{\hat h}_{ab}|} \, d\psi \wedge d \theta_1 \wedge d\phi_1 \wedge d\theta_2 \wedge d\phi_2$ and where ${\hat h}_{ab}$ is  defined by $d\Sigma_{1,1}^2 = {\hat h}_{ab} dy^a dy^b$. From (4.8) and (4.17) one finds
\begin{equation}
\frac{1}{2\kappa_{10}^2} \int {}^{*}F_5 \,=\,\mu_{D} \,\,,
\end{equation}
which confirms the fact that $\mu_{D}$ is an R-R electric charge carried by a D3-brane located at $\vec{\hat r} =0$. $\mu_{D}$, on the other hand, becomes a magnetic charge in the dual theory in which ${\mathcal F}_5$ is given in terms of ${}^{\ast} F_5$. The magnetic charge associated with ${}^{\ast} F_5$ of the dual theory is also defined by (4.18), and therefore $\mu_{D}$ becomes a self-dual charge $-$ the charge being both electric and at the same time magnetic $-$ once the self-duality condition is imposed.

Let us introduce a new coordinate $r$ defined by $d{\hat r}/R^5 = dr/r^5$. In terms of the coordinate $r$ the six-dimensional metric (4.4) becomes
\begin{equation}
ds_6^2 = \Big(\frac{R}{r}\Big)^{10} \Big[ dr^2 + \Big(\frac{r}{R} \Big)^8 r^2 \,d \Sigma_{1,1}^2 \Big] \,\,,
\end{equation}
which reduces to the conifold metric (2.2) when $R=r$. Using (4.17) one finds that the set of four linearly independent field equations (4.12)-(4.15) can be rewritten in the covariant form as
\begin{equation}
\nabla^2 \ln R - \frac{4}{R^2} \Big( \frac{R}{r}\Big)^{10} - \frac{1}{4}\,e^{\hat{\Phi} -2B}\Big(\frac{R}{r}\Big)^{10} \beta=0\,\,,
\end{equation}
\begin{equation}
\nabla^2 \hat{\Phi} = 2 \kappa_{10}^2 \, g_s^2 \, \Big[ e^{2B} \,  \Big(T(\Phi) + \frac{\partial T(\Phi)}{\partial \Phi} \Big) - \frac{\partial \mu (\Phi)}{\partial \Phi} \xi(r) \Big] \,\delta^6 ({\vec r}\,)\,\,,
\end{equation}
\begin{eqnarray}
\nabla^2 B - \frac{1}{2}\,e^{-2\hat{\Phi} +4B}\,\, \frac{q_{D}^2}{r^{10}} - \frac{1}{2}\,e^{\hat{\Phi} -2B}\Big(\frac{R}{r}\Big)^{10} \beta ~~~~~~~~~~~~~~~~~~~~\nonumber \\ ~~~~~~~~~~\,\,= 2\kappa_{10}^2 \,{g_s^2}\,\Big[ e^{2B} \,\Big( T(\Phi) + \frac{1}{2}\frac{\partial T(\Phi)}{\partial \Phi}\Big) -\frac{1}{2} \frac{\partial \mu (\Phi)}{\partial \Phi} \xi(r) \Big] \,\delta^6 ({\vec r}\,) \,\,,
\end{eqnarray}
\begin{equation}
-\mathcal{H} + \frac{1}{4}\,e^{-2\hat{\Phi} +4B}\,\, \frac{q_{D}^2}{r^{10}} - \frac{1}{2}\,e^{\hat{\Phi} -2B}\Big(\frac{R}{r}\Big)^{10} \beta + \frac{10}{R^2} \Big[ {R^{\prime}}^2 - \Big(\frac{R}{r}\Big)^{10} \, \Big] =0 \,\,,
\end{equation}
where ${q_{D}} = 2\kappa_{10}^2 \, g_s \, \mu_{D} /{\rm Vol(B)}$, and $\nabla^2$ is the Laplacian $\nabla^2 = (1/r^5 )(d/dr)(r^5 d/dr)$ defined on the conifold (so $\delta^6 ({\vec r}\,)$ is now normalized by $\int r^5 dr \, \epsilon_5 \, \delta^6({\vec r}\,) =1$). Also $\mathcal{H}$ is defined as before, but now the prime denotes the derivative with respect to $r$ instead of $\hat r$. Note that (4.20) has been obtained by adding (4.12) and (4.13), and (4.23) is a rewrite of (4.12).

Now we turn to the four-dimensional field equation, which is just the Einstein equation defined on the four-dimensional spacetime. In order to find them we rewrite the total action in the form
\begin{equation}
I_{\rm total} = \frac{1}{2 \kappa^2} \Big[ \int d^4 x \sqrt{-g_4}\, {\mathcal R}_4 (g_{\mu\nu})  - 2 \lambda \int d^4 x \sqrt{-g_4} \Big]\,\,,
\end{equation}
where $\lambda$ is defined by
\begin{equation}
\lambda = - \kappa^2 \big[ {\hat I}_{\rm bulk} + {\hat I}_{\rm brane} \big] \,\,,
\end{equation}
where
\begin{eqnarray}
{\hat I}_{\rm bulk} = \frac{1}{2 \kappa_{10}^2 \, g_s^2}  \int d^6 y \sqrt{h_6}\,\,\big[{\mathcal R}_6 (h_{mn}) - (\partial \hat{\Phi})^2 + 2(\partial \hat{\Phi})(\partial B) \nonumber \\ - 2(\partial B)^2 +\frac{g_s^2}{2} \, e^{2\hat{\Phi} - 4B} (\partial \xi)^2 \big]\,\,,~~~~~
\end{eqnarray}
and ${\hat I}_{\rm brane} = I_{\rm brane} / \big[ \int d^4 x \sqrt{-g_4} \,\big]$, which is the brane action per unit volume of the four-dimensional spacetime. Also in (4.24), $2\kappa^2$ is defined by
\begin{equation}
\frac{1}{2\kappa^2} = \frac{1}{2 \kappa_{10}^2 \, g_s^2} {\int d^6 y \sqrt{h_6} \,\,e^{\hat{\Phi} - 2B} } \,\,,
\end{equation}
which is identified as an inverse square of the four-dimensional Planck mass, $2 \kappa^2 =1 / M_{pl}^2$. Since (4.24) yields the Einstein equation ${\mathcal R}_{\mu\nu} -\frac{1}{2} g_{\mu\nu} {\mathcal R}_4 + \lambda g_{\mu\nu} =0$, the above $\lambda$ is identified as the four-dimensional cosmological constant, and one finds that $\beta = 4 \lambda$ from the definition of $\beta$.

To evaluate $\lambda$, substitute (4.20) and (4.23) into (4.26) and use
\begin{equation}
\int d^6 y \sqrt{h_6} \, {\mathcal R}_6 (h_{mn}) \,=\, \int r^5 dr \epsilon_5 \, \Big[-10 \,\nabla^2 \ln R \,+\, \frac{20}{R^2} \big[{R^{\prime}}^2 + \Big(\frac{R}{r}\Big)^{10} \,\big] \Big] \,\,,
\end{equation}
together with
\begin{equation}
\partial_r \xi = \frac{q_{D}}{g_s} \,\frac{e^{4B -2\hat{\Phi}}}{r^5} \,\,,
\end{equation}
and (4.27). Then we obtain
\begin{equation}
{\hat I}_{\rm bulk} = - {3 \beta}/{4 \kappa^2} \,\,,
\end{equation}
and from (4.25),
\begin{equation}
\lambda = \frac{\kappa^2}{2} {\hat I}_{\rm brane} \,\,,
\end{equation}
where we have used the relation $\beta = 4 \lambda$. (4.31) shows that $\lambda$ is just (the value of) the brane action density times the inverse square of the Planck mass $M_{pl}$. The brane action, which is a world-volume action of the D-brane, consists of two terms. (4.11) shows that the first term ($\equiv I_{\rm brane}^{(\rm NS)}$) is the DBI action representing the coupling of D-brane to the closed string fields $G_{\mu\nu}$ and $\Phi$ of the NS-NS sector. The second term ($\equiv I_{\rm brane}^{(\rm R)}$) is an R-R counterpart of the first term since it represents a coupling of D-brane to the R-R four-form $A_4$. We need this term if the D3-brane is a BPS state.

$I_{\rm brane}^{(\rm NS)}$ contains the tension of the D3-brane $T(\Phi) =T_0 \,e^{-\Phi} + \rho_{\rm vac} (\Phi)$, where $\rho_{\rm vac}(\Phi)$ takes the form $\rho_{\rm vac}(\Phi)= \sum_{n=0}^{\infty} \rho_n \, e^{n \Phi}$. Then from (4.11) and (4.31) (and assuming that $e^{2B} \sim O(1)$), the value of $\lambda$ arising from $\rho_{\rm vac}$ in $I_{\rm brane}^{(\rm NS)}$ is expected to be $\sim \mu^4 / M_{pl}^2$, where $\mu$ is the large-momentum cutoff at which the supersymmetry is broken. So if we take $\mu \sim TeV$ (also see footnote 4), $\mu^4 / M_{pl}^2$ will be of an order $(TeV)^4 / (10^{19}GeV)^2$, which is too large to be a correct value for the present $\lambda$. In our case, however, there is another contribution to $\lambda$ coming from $I_{\rm brane}^{(\rm R)}$ which is also expected to be of the same order as that of the first term. So we expect that the near-vanishing $\lambda$ can be obtained if these two terms cancel each other. In the next section we will show that this is indeed the case.

\vskip 1cm
\hspace{-0.65cm}{\bf \Large {V. Vanishing $\bf \lambda$ and the supersymmetry of the bulk $~~~~\,$ region}}
\vskip 0.5cm
\setcounter{equation}{0}
\renewcommand{\theequation}{5.\arabic{equation}}

In the previous section we have found that the four-dimensional cosmological constant $\lambda$ is given by $I_{\rm brane}$ times a constant $(1/M_{pl}^2 )[\int d^4 x \sqrt{-g_4}\,]^{-1}$. But $I_{\rm brane}$ consists of two parts, $I_{\rm brane}^{\rm (NS)}$ and $I_{\rm brane}^{\rm (R)}$, which suggests that the vacuum energies on the brane must also appear in two types, NS-NS type and R-R type. In this section we will show that $\lambda$ is forced to vanish by field equations and the vanishment of $\lambda$ requires that the two types of vacuum energies on the brane cancel to all orders of $g_s$. Also the cancelation between NS-NS and R-R vacuum energies leads to a constant dilaton, $\Phi =0$, in the bulk region. So the supersymmetry is not broken in the bulk region even under vacuum fluctuations on the brane.

\vskip 0.3cm
\hspace{-0.6cm}{\bf (A) case I}

To start the discussion let us go back to the field equations (4.20)-(4.23), and consider the simplest situation in which we have only a background conifold geometry. Since D3-brane has not been introduced yet, the tension and charge both vanish: $T(\Phi) = \mu(\Phi) =0$ ($q_{D} = 0$), and the field equations are trivially solved by\footnote{Note that (5.1) is the only solution to (4.20)-(4.23) which accords with an assumption that the background internal space around D3-brane is a conifold ($R(r)=r$) of the Calabi-Yau threefold. Indeed in Sec. 8.3 it is precisely shown that for the given action (8.14) $\beta$ must satisfy $\beta=0$. Since (4.10) is a simple case of (8.14) the corresponding solution of (4.10) must also satisfy $\beta=0$. So (5.1) is the only correct solution to the given field equations.}
\begin{equation}
R(r) =r \,\,,~~~\hat{\Phi} (r)= B(r)=0\,\,, ~~~\beta=0 \,\,.
\end{equation}
So from (4.3) and (4.19) (also see (4.9)) one obtains $ds_{10}^2 =ds_{0123}^2 + ds_{\rm conifold}^2$ with $\lambda=0$, which coincides with the given configuration as it should be.

The configuration of the case I preserves some unbroken supersymmetry because the dilaton $\Phi$ (as well as other fields) vanishes there. In ten-dimensions the supersymmetry variation of the dilatino always contains a first derivative of $\Phi$. In the absence of two-form gauge fields (and in the absence of R-R zero-form for the type IIB)\footnote{This, together with $H_3 =0$, precisely coincides with our configuration under discussion.} the variations of the fermion fields are given, to a leading order in $g_s$, by\footnote{In the type I the terms with $H_{mnp}$ in both $\delta \psi_m$ and $\delta \chi_{\phi}$ are absent. And in the heterotic type the $H_{mnp}$ term in $\delta \psi_m$ appears with the spin connection term which however vanishes as $\alpha^{\prime} \rightarrow 0$.}
\begin{equation}
\delta \psi_m = D_m \eta \mp \frac{1}{8} \Gamma^{np}H_{mnp} \eta\,\,, ~~~~~~ \delta \chi_{\phi} = \big( \pm\Gamma^{m} \partial_{m} \Phi - \frac{1}{12} \Gamma^{mnp} H_{mnp} \big)\, \eta \,\,,
\end{equation}
where $\psi_M$ and $\chi_{\phi}$ are gravitino and dilatino, respectively. Thus in the absence of $H_3$ the supersymmetry is unbroken when $\Phi$ is constant.\footnote{It is well known \cite{17} that in the type IIB supergravity the supersymmetry requires $A=-B$, and therefore ${\hat \Phi}=0$ from (4.9).} As a result the case I preserves $1/4$ supersymmetry because the background is compactified on a Calabi-Yau. Thus for instance in the type IIB the unbroken supersymmetry of the case I will be ${\mathcal N}=2$, $d=4$ after reduction.

\vskip 0.3cm
\hspace{-0.6cm}{\bf (B) case II}

Next we introduce a D3-brane at the conifold singularity, but the quantum corrections (the vacuum energy) are still neglected at this point. In the case II the tension and charge are respectively given by $T(\Phi) =T_0 e^{-\Phi}$ and $\mu (\Phi)= \mu_0$, and therefore $q_{D}=q_0$ where $q_0 = 2 \kappa_{10}^2 g_s \mu_0 /{\rm Vol(B)}$. The solution to (4.20) and (4.21) however takes the same form as that of the case I because (4.20) does not include any $T(\Phi)$ or $\mu (\Phi)$ (and $q_{D}$) term, and the right-hand side of (4.21) vanishes for the given $T(\Phi)$ and $\mu (\Phi)$. So except for $B(r)$, the solution $R(r) =r,~\hat{\Phi} (r)= 0$ and $\beta=0$ of the case I will still be the right solution for the case II either as long as the equations
\begin{equation}
\nabla^2 \ln \chi - 2 \, \frac{{q_0^2}}{r^{10}} \, \chi= 2 c_B \,\delta^6 ({\vec r}\,)\,\,,
\end{equation}
and
\begin{equation}
\Big(\frac{d}{dr} \ln \chi \Big)^2 -4 \,\frac{{q_0^2}}{r^{10}} \, \chi = 0 \,\,,
\end{equation}
which follow (after setting $R(r)=r$, $\hat{\Phi}(r)=0$ and $\beta =0$) from the remaining equations (4.22) and (4.23), admit a nonsingular solution. In (5.3) and (5.4), $\chi =e^{4B -2 \hat{\Phi}}$ (with $\hat{\Phi}=0$) and
\begin{equation}
c_B = 2 \kappa_{10}^2 \, g_s \, \chi^{1/2} (0) \, T_0 \,\,.
\end{equation}

The solution satisfying both (5.3) and (5.4) can be obtained as follows. The solution to (5.3) takes the form \cite{19}
\begin{equation}
\chi(r) = \frac{a_0}{[X-X^{-1}]^2}\,\,, ~~~~~\Big( X= e^{-({\alpha}/{4})[c_0 + ({r_0}/{r})^4]} \Big) \,\,,
\end{equation}
where $c_0$ and $r_0$ are constants, and (assuming that $\alpha >0 $)
\begin{equation}
a_0 = \frac{4\,\alpha^2 \,r_0^8}{q_0^2}\,\,, ~~~~~\alpha =\frac{c_B}{r_0^4 \,{\rm Vol(B)}}\,\,.
\end{equation}
But since $\chi(r) \simeq a_0 X^2 \simeq a_0 \exp[-(\alpha/2)(r_0 /r)^4]$ as $r \rightarrow 0$, $\chi(r)$ goes to zero as $r \rightarrow 0$, and from (5.5) one finds that $c_B$ and consequently $\alpha$ both vanish. Since $\alpha$ vanishes, (5.6) reduces to
\begin{equation}
\chi(r)=\Big( 1+\frac{\rm Q_0}{r^4} \Big)^{-2} \equiv H_{\rm D}^{-2}\,\,, ~~~~~~~~~~\Big( {\rm Q_0} \equiv \frac{q_0}{4}\Big)\,\,,
\end{equation}
where $c_0$ and $r_0$ have been so adjusted as to satisfy the asymptotic requirement $\chi(r) \rightarrow 1$ as $r \rightarrow \infty$. Note that (5.8) still satisfies the boundary condition $\chi(r) \rightarrow 0$ as $r \rightarrow 0$.

(5.8) is obviously the harmonic function for the D3-brane since $\rm Q_0$ is proportional to $g_s$. Indeed it coincides with the usual harmonic function for the D3-brane if we take ${\rm Q_0} = g_s {\alpha^{\prime}}^2$. Also one can check that (5.8) satisfies the remaining equation (5.4) as well. After all, the solution for the case II, which is a classical (on-shell) solution for the configuration of D3-brane located at the conifold singularity, is summarized as
\begin{equation}
R(r) =r \,\,,~~~~e^{\hat{\Phi} (r)}= 1\,\,,~~~~ e^{B(r)}=\Big( 1+ \frac{\rm Q_0}{r^4} \Big)^{-1/2}\,\,, ~~~~\beta=0 \,\,.
\end{equation}
The metric is therefore written as
\begin{equation}
ds_{10}^2 = H_{\rm D}^{-1/2}\,\big[-dt^2 + d \vec{x}_3^2 \,\big] +  H_{\rm D}^{1/2}\,\big[dr^2 +r^2 d \Sigma_{1,1}^2 \,\big] \,\,,
\end{equation}
which is the usual D3-brane metric and coincides with (2.13) of Sec. II as well.

Also since $\beta$ vanishes the two terms $I_{\rm brane}^{\rm (NS)}$ and $I_{\rm brane}^{\rm (R)}$ must cancel to satisfy $\lambda=0$. This is indeed the case and we can show it as follows. For $T(\Phi) = T_0 \,e^{-\Phi}$ and $\mu (\Phi) = \mu_0$, $I_{\rm brane}$ is simply
\begin{equation}
I_{\rm brane}= \big[\int d^4 x \, \sqrt{-g_4} \, \big] \int r^5 dr \epsilon_5 \, \Big[ - \frac{1}{g_s} \chi^{1/2} T_0 + \mu_0 \,\xi \,\Big]\, \delta^6 (\vec{r}) \,\,.
\end{equation}
But using (4.29) and (5.9) one finds that
\begin{equation}
\xi (r) = \frac{1}{g_s} \Big(1+ \frac{Q_0}{r^4} \Big)^{-1} \,=\, \frac{1}{g_s}\, \chi^{1/2}(r) \,\,.
\end{equation}
So (5.11) vanishes if $T_0 = \mu_0$. Namely $\lambda =0$ is automatically satisfied if the D3-brane is a BPS state. This agrees with the requirement that the nonzero cosmological constant on the D-brane should arise from the quantum fluctuations, not from the D-brane itself. Since we have ignored the quantum corrections in the case II, $\lambda$ and consequently $\beta$ must vanish as in (5.9).

The case II preserves some unbroken supersymmetry because the dilaton is constant, and it would be half of that of the case I because the D3-brane will reduce the supersymmetry into a half. Thus the unbroken supersymmetry expected for the type IIB will be ${\mathcal N}=1$, $d=4$ after reduction.

\vskip 0.3cm
\hspace{-0.6cm}{\bf (C) case III}

In the case III we still consider the configuration of D3-brane located at the conifold singularity, but now the tension and charge of the D3-brane include quantum correction terms: $T(\Phi) = T_0 e^{-\Phi} + \rho_{\rm vac}(\Phi)$ and $\mu (\Phi) = \mu_0 + \delta \mu (\Phi)$. Since sum of two terms of $I_{\rm brane}$ vanishes at the tree level of the D3-brane (case II), the substantial contribution to $\lambda$ must come from the quantum correction terms:
\begin{equation}
\lambda = \frac{\kappa^2}{2} \big[\delta_Q I_{\rm brane}^{\rm (NS)} + \delta_Q I_{\rm brane}^{\rm (R)} \big]/\big[ \int d^4 x \sqrt{-g_4} \big]\,\,.
\end{equation}
Though $\delta_Q I_{\rm brane}^{\rm (NS)}$ and $\delta_Q I_{\rm brane}^{\rm (R)}$ in (5.13) are caused by the quantum fluctuations, they themselves are classical (on-shell) quantities just like the vacuum energy densities $\rho_{\rm vac}$ and $\delta \mu$. $\rho_{\rm vac}$ and $\delta \mu$ are both macroscopic quantities obtained by integrating local fluctuations all over the four-dimensional spacetime. So they are somethings like spacetime averages of the local fluctuations which are microscopic quantum phenomena. Since they are all on-shell quantities they must satisfy the field equations through $\lambda$. As mentioned in Sec. IV the change in $\lambda$ caused by $\rho_{\rm vac}$ in $\delta_Q I_{\rm brane}^{\rm (NS)}$ is expected to be of an order $\sim (TeV)^4 / (10^{19}GeV)^2$, and this must be canceled anyway by the change of the second term, $\delta_Q I_{\rm brane}^{\rm (R)}$, in order to obtain near-vanishing $\lambda$ (more precisely it is $\sim ({\rm milli}$-$eV)^4 /(10^{19}GeV)^2$). $\delta_Q I_{\rm brane}^{\rm (R)}$ due to quantum fluctuations really occurs as follows.

At the classical level $I_{\rm brane}^{\rm (R)}$ is given by an electric coupling of the R-R four-form to the four-dimensional world volume of the D3-brane: $I_{\rm brane}^{\rm (R)} = \mu_0 \int A_4$. But this can be rewritten as
\begin{equation}
I_{\rm brane}^{\rm (R)} = \frac{1}{4!}\int d^4 x \, A_{\mu_0 \mu_1 \mu_2 \mu_3}\,J^{\mu_0 \mu_1 \mu_2 \mu_3} \,\,,
\end{equation}
where $J^{\mu_0 \mu_1 \mu_2 \mu_3}$ is the world volume current density of the D3-brane:
\begin{equation}
J^{\mu_0 \mu_1 \mu_2 \mu_3} = \mu_0 \, \epsilon^{\alpha_0 \alpha_1 \alpha_2 \alpha_3} \big(\frac{\partial X^{\mu_0}} {\partial x^{\alpha_0}} \big) \cdots \big(\frac{\partial X^{\mu_3}}{\partial x^{\alpha_3}} \big) \,\,.
\end{equation}
At the classical level $J^{\mu_0 \mu_1 \mu_2 \mu_3}$ is just a solitonic current density, $J_{\rm sol}^{\mu_0 \mu_1 \mu_2 \mu_3}$, representing classical world volume dynamics of D3-brane. In that case $X^{\mu}(x)$'s in (5.15) stand for classical fields, $X_{\rm cl}^{\mu}(x)$, defined on the world volume of the D3-brane, and for the embedding $X_{\rm cl}^{\mu}(x)=x^{\mu}$, $J_{\rm sol}^{0123}$ is simply $\mu_0$. At the quantum level, however, $X^{\mu}(x)$'s include fluctuations and we have to separate each of them into a classical part and the fluctuation, $X^{\mu} = X_{\rm cl}^{\mu} + {X^{\mu}}^{\prime}$. By this separation $J^{\mu_0 \mu_1 \mu_2 \mu_3}$ can be written as $J^{\mu_0 \mu_1 \mu_2 \mu_3} = J_{\rm sol}^{\mu_0 \mu_1 \mu_2 \mu_3} +\, <\chi_{\rm vac}^{\mu_0 \mu_1 \mu_2 \mu_3}>$ where $J_{\rm sol}^{\mu_0 \mu_1 \mu_2 \mu_3}$ is the classical current density as mentioned above, while $<\chi_{\rm vac}^{\mu_0 \mu_1 \mu_2 \mu_3}>$ is an R-R counterpart of $\rho_{\rm vac}$ representing quantum corrections arising from the fluctuations on the D3-brane. Finally denoting $J_{\rm sol}^{0123}$ and $<\chi_{\rm vac}^{0123}>$ by $\mu_0$ and $\delta \mu (\Phi)$, respectively, one can rewrite (5.14) as
\begin{equation}
I_{\rm brane}^{\rm (R)} = \big[\int d^4 x \, \sqrt{-g_4} \,\big] \int r^5 dr \epsilon_5 \,  \mu (\Phi) \, \xi(r) \delta^6 (\vec{r}) \,\,,
\end{equation}
which is just the second term of (4.11) and where $\mu (\Phi) = \mu_0 + \delta \mu (\Phi)$. (5.16) shows that $\delta_Q I_{\rm brane}^{\rm (R)}$ is due to $\delta \mu(\Phi)$, the R-R counterpart of $\rho_{\rm vac}$.

Let us go back to the field equations (4.20)-(4.23), where $T(\Phi)$ and $\mu(\Phi)$ are now given by $T(\Phi)= T_0 \, e^{-\Phi} +\rho_{\rm vac}$ and $\mu(\Phi)= \mu_0 + \delta \mu (\Phi)$. The field equation (4.20) does not change and still can be solved by $R(r)=r$ and $\beta=0$. So $\lambda$ must vanish by (4.20). Upon setting $R(r)=r$ and $\beta=0$ the remaining field equations can be recast into
\begin{equation}
\nabla^2 \hat{\Phi} = \delta c_{\Phi} \, \delta^6 (\vec{r})\,\,,
\end{equation}
\begin{equation}
\nabla^2 \ln \chi -2 \,\frac{q_{\rm D}^2}{r^{10}}\,\chi= 2(c_B + \delta c_B )\, \delta^6 (\vec{r})\,\,,
\end{equation}
\begin{equation}
\Big( \frac{d}{dr} \ln \chi \Big)^2 - 4 \,\frac{q_{\rm D}^2}{r^{10}}\,\chi + 4 \Big(\frac{d \hat{\Phi}}{dr} \Big)^2 =0 \,\,,
\end{equation}
where $\chi=e^{4B-2\hat{\Phi}}$ and $c_B = 2 \kappa_{10}^2 \, g_s \, \chi^{1/2} (0) \, T_0$ as before but $\delta c_B$ and $\delta c_{\Phi}$ are
\begin{equation}
\delta c_B = 2 \kappa_{10}^2 \, g_s \, \chi^{1/2} (0) \, e^{\Phi}\rho_{\rm vac}\big|_{\Phi=\Phi(0)}\,\,,\nonumber
\end{equation}
\begin{equation}
\delta c_{\Phi}= 2 \kappa_{10}^2 \, g_s^2 \Big[\,\frac{1}{g_s} \chi^{1/2}(0) \,e^{\Phi} \big( \rho_{\rm vac}+ \frac{\partial \rho_{\rm vac}}{\partial \Phi} \big) - \frac{\partial \delta \mu}{\partial \Phi} \, \xi(0)\,\Big]_{\Phi=\Phi(0)}\,\,.
\end{equation}
Equations (5.18) and (5.19) have the same form as (5.3) and (5.4) except that the constants $q_0$ and $c_B$ are replaced by $q_{\rm D}$ and $c_B + \delta c_B$. Using $c_B + \delta c_B \rightarrow 0$, one obtains
\begin{equation}
\chi(r)=\Big( 1+\frac{\rm Q_D}{r^4} \Big)^{-2}\,\,, ~~~~~~~~~~\Big( {\rm Q_D} \equiv \frac{q_D}{4}\Big)\,\,,
\end{equation}
and from (4.29)
\begin{equation}
\xi(r)=\frac{1}{g_s} \Big( 1+\frac{\rm Q_D}{r^4} \Big)^{-1} = \frac{1}{g_s} \chi^{1/2}(r)\,\,.
\end{equation}

We have just seen that the condition $\lambda =0$ is still maintained in case III as well. This means that the field equation (4.20) forces $\delta_Q I_{\rm brane}^{\rm (NS)}$ and $\delta_Q I_{\rm brane}^{\rm (R)}$ to adjust themselves to cancel so that $\lambda$ vanishes even at the quantum level. Omitting the tree level terms in (5.11) one finds that $I_{\rm brane}$, which now consists of $\delta_Q I_{\rm brane}^{\rm (NS)}$ and $\delta_Q I_{\rm brane}^{\rm (R)}$, is given by
\begin{equation}
I_{\rm brane} = \big[\int d^4 x \, \sqrt{-g_4} \,\big] \int r^5 dr \epsilon_5 \, \Big[\,-\frac{1}{g_s} \chi^{1/2} e^{\Phi} \, \rho_{\rm vac}+ \delta \mu \, \xi \,\Big]\, \delta^6 (\vec{r})\,\,.
\end{equation}
So the condition $\lambda =0$ requires that
\begin{equation}
 \frac{1}{g_s} \chi^{1/2} e^{\Phi} \, \rho_{\rm vac} =\delta \mu \, \xi \,\,,
\end{equation}
which in turn implies
\begin{equation}
e^{\Phi} \, \rho_{\rm vac} =\delta \mu
\end{equation}
by (5.22). Further, using the perturbative expansions $\rho_{\rm vac}= \sum_{n=0}^{\infty} \rho_n e^{n\Phi}$ and $\delta \mu= \sum_{n=1}^{\infty} \mu_n e^{n\Phi}$ one can express (5.25) in terms of the coefficients $\rho_n$ and $\mu_n$ as
\begin{equation}
\rho_n = \mu_{n+1} \,\,,
\end{equation}
where $n$ represents non-negative integers. (5.26) is a result of $\lambda=0$ and it shows that the two types of vacuum energies on the brane must cancel to all orders of $g_s$.

It is well known \cite{20} that (5.26) is really satisfied for $n=0$, where $\rho_0$ and $\mu_1$ are identified as the NS-NS and R-R sector one-loop amplitudes of the unoriented string theory: $\rho_0 \sim i \mathcal{A}_{\rm NS}/V_{10}$ and $\mu_1 \sim - i \mathcal{A}_{\rm R}/V_{10}$ with $\mathcal{A}$ ($\equiv \mathcal{A}_{\rm NS} + \mathcal{A}_{\rm R}$) given by
\begin{equation}
\mathcal{A}= \int_{T_2} \frac{d^2 \tau}{4 \tau_2} \, {\rm{Tr}}_{\rm{NS+R}} \Bigg\{ \Omega_{(F, \tilde{F})}\, q^{L_0} \bar{q}^{\bar{L}_0} \Bigg\}
\end{equation}
for the closed string, and similarly for the open string. In (5.27), $q=\exp(2 \pi i \tau)$ and $\Omega_{(F, \tilde{F})}$ is an appropriate GSO projection of the given theory. For instance for type IIB it will be $\Omega_{(F, \tilde{F})} = \frac{1+(-1)^F}{2} \frac{1+(-1)^{\tilde F}}{2}$. The amplitude $\mathcal{A}$ vanishes for the BPS states and the cancelation between two sectors is achieved by Jacobi's abstruse identity. In fact (5.26) (and equivalently (5.25)) can be regarded as a restatement of the BPS condition, and consequently some spacetime supersymmetry is still expected in the case III. Note that $\delta c_{\Phi}$ in (5.20) vanishes by (5.24) because the right-hand side of (5.20) is just a functional $\Phi$ derivative of (5.24), and therefore we have $\hat{\Phi}=0$ from (5.17) as in the case II. The unbroken supersymmetry of the case II is still preserved in the case III as well. The vacuum fluctuations on the D-brane do not break the supersymmetry of the bulk region.

\vskip 1cm
\hspace{-0.65cm}{\bf \Large {VI. Supersymmetry breaking}}
\vskip 0.5cm
\setcounter{equation}{0}
\renewcommand{\theequation}{6.\arabic{equation}}

In the previous section we have seen that vacuum fluctuations on the BPS D3-brane do not break the supersymmetry of the bulk region because the dilaton remains a constant there even under fluctuations on the D3-brane. Such a result seems to be natural for the BPS state since in that case the right-hand side of the field equation for $\Phi$ vanishes by the cancelation between NS-NS and R-R sector vacuum energies on the brane. So we have a constant $\Phi$ and hence an unbroken supersymmetry. This, however, is not to be the case anymore when we go to the brane region. In this section we will show that the $d=4$ supersymmetry is essentially broken in the brane region due to the gauge symmetry breaking of the R-R four-form arising at the quantum level.

\vskip 0.5cm
\hspace{-0.65cm}{\bf \large 6.1 Gauge symmetry breaking induced by quantum fluctuations}
\vskip 0.2cm

The total action (4.1) plus (4.2) remains invariant under the gauge transformation $A_4 \rightarrow A_4 + \delta A_4$ with $\delta A_4 = d \Lambda_3$ where $\Lambda_3$ is an arbitrary three-form. $I_{\rm bulk}$ is invariant because so is $F_5$ in $I_{\rm bulk}$. $I_{\rm brane}$ is also invariant because the variation $\delta I_{\rm brane}^{(R)}$ vanishes for $\delta A_4 = d \Lambda_3$: $\delta I_{\rm brane}^{(R)}=\mu_0 \int_{\partial \Sigma} \Lambda_3 =0$ where $\partial \Sigma$ is the boundary of the four-dimensional spacetime.\footnote{Here we assumed that the four-dimensional spacetime has no boundary.} But this is valid only at the classical level. When we go up to quantum level $I_{\rm brane}^{(R)}$ is not gauge invariant anymore though $I_{\rm bulk}$ still remains gauge invariant. Note that $I_{\rm brane}^{(R)}$ can be written as (5.14) and where $J^{\mu_0 \mu_1 \mu_2 \mu_3}$ consists of two parts, $J_{\rm sol}^{\mu_0 \mu_1 \mu_2 \mu_3}$ and $<\chi_{\rm vac}^{\mu_0 \mu_1 \mu_2 \mu_3}>$. The current density $J_{\rm sol}^{\mu_0 \mu_1 \mu_2 \mu_3}$ satisfies $\partial _{\mu_0}J_{\rm sol}^{\mu_0 \mu_1 \mu_2 \mu_3}=0$ because it is a tree level (on-shell) quantity. But $\chi_{\rm vac}^{\mu_0 \mu_1 \mu_2 \mu_3}$ does not necessarily satisfy $<\partial_{\mu_0} \chi_{\rm vac}^{\mu_0 \mu_1 \mu_2 \mu_3}> =0$ because it is an off-shell quantity arising from the quantum fluctuations. Thus the total $J^{\mu_0 \mu_1 \mu_2 \mu_3}$ is not locally conserved at the quantum level and the gauge transformation $\delta A_{\mu_0 \mu_1 \mu_2 \mu_3}= 4 \,\partial_{[\mu_0} \,\Lambda_{\mu_1 \mu_2 \mu_3]}$ induces a nonzero variation of $I_{\rm brane}^{\rm (R)}$. Integrating by parts one obtains
\begin{equation}
\delta I_{\rm brane}^{\rm (R)} =-\,\frac{1}{3!}\int d^4 x \, \Lambda_{\mu_1 \mu_2 \mu_3} <\partial_{\mu_0} \chi_{\rm vac}^{\mu_0 \mu_1 \mu_2 \mu_3}> \,\,,
\end{equation}
and since $<\partial_{\mu_0} \chi_{\rm vac}^{\mu_0 \mu_1 \mu_2 \mu_3}>$ is nonzero (6.1) does not generally vanish at the quantum level. $\delta I_{\rm brane}^{(R)}$, however, vanishes if $\Phi$ (and therefore $\delta \mu (\Phi)$) is independent of $x$. In terms of $\delta \mu (\Phi)$ $\delta I_{\rm brane}^{(R)}$ can be rewritten as
\begin{equation}
\delta I_{\rm brane}^{\rm (R)}=\int d^4 x \, \delta \mu (\Phi) \,\partial_{[0}\Lambda_{123]} \,\,,
\end{equation}
and (6.2) vanishes if $\delta \mu (\Phi)$ is independent of $x$ because in that case the integrand becomes a total derivative in $x$.

In addition to (6.2) there is another important variation of $I_{\rm brane}^{\rm (R)}$ which plays a crucial role in addressing the cosmological constant problem. To find its explicit form rewrite it as $\delta I_{\rm brane}^{\rm (R)} = \mu_0 \int d \Lambda_3$ and take an ansatz
\begin{equation}
\Lambda_3 = F(y)\,\sqrt{-g_4} \, dx^1 \wedge dx^2 \wedge dx^3 \,\,,
\end{equation}
where $F(y)$ is an arbitrary function of the internal coordinates $y^{m}$. (6.3) is the most appropriate ansatz for $\Lambda_3$, which accords with (4.5). Taking a derivative to $\Lambda_3$ one obtains
\begin{equation}
\delta I_{\rm brane}^{\rm (R)} =\int d^4 x \, \sqrt{-g_4}\, f_m (y) J^{m123} \,+\, \frac{3}{2} \int d^4 x \, \sqrt{-g_4}\, H\, F(y) J^{0123} \,\,,
\end{equation}
where $f_m (y)$ ($\equiv \partial_m F(y)$) represents $\delta A_{m123}/\sqrt{-g_4}$, and $H$$\big($$\equiv$$(2/3)\, \partial_0 \ln \sqrt{-g_4} \big)$ is Hubble constant of the four-dimensional spacetime $ds_4^2 = -dt^2 + e^{Ht} d \vec{x}_3$, which therefore vanishes for $\lambda =0$ because $\lambda \propto H^2$. Also in (6.4) $J^{m123}$ is defined by
\begin{equation}
J^{m123} = \mu_0 \epsilon^{\alpha_0 \alpha_1 \alpha_2 \alpha_3}\, \big( \frac{\partial Y^m}{\partial x^{\alpha_0}}\big)\wedge \big( \frac{\partial X^1}{\partial x^{\alpha_1}}\big) \wedge \big( \frac{\partial X^2}{\partial x^{\alpha_2}}\big)\wedge \big( \frac{\partial X^3}{\partial x^{\alpha_3}}\big) \,\,,
\end{equation}
as in $J^{0123}$.

At the classical level $J^{m123}$ is simply $J_{\rm sol}^{m123}$ where $Y^m$ and $X^{\mu}$ are classical fields, and it vanishes for the embedding $X^{\mu}(x) = x^{\mu}$ because ${\partial Y^m}/{\partial x^{\alpha_0}}=0$. So the nonvanishing contribution to $J^{m123}$ comes from the quantum excitations
$<\chi_{\rm vac}^{m123}>$. Denoting $<\chi_{\rm vac}^{m123}>$ by $\delta \mu_{\rm T}^m (\Phi)$, one can rewrite (6.4) as
\begin{equation}
\delta I_{\rm brane}^{\rm (R)} =\big[\int d^4 x \, \sqrt{-g_4} \big]\int r^5 dr \epsilon_5 \, \delta \mu_{\rm T}^m (\Phi)\, f_m (y)\, \delta^6 (\vec{r})\,\,,
\end{equation}
where we have omitted the second term since we always consider the configuration with $\lambda =0$ as required by (4.20). By (6.6) the field equations necessarily acquire extra terms. But in effect only (4.21) among (4.20)-(4.23) is revised by the additional contribution $\delta I_{\rm brane}^{\rm (R)}$. The other equations (when expressed in terms of $\chi(= e^{4B- 2 \hat{\Phi}})$) remain unchanged as one can check from the total brane action $I_{\rm brane} + \delta I_{\rm brane}^{\rm (R)}$. The revised field equation is
\begin{equation}
\nabla^2 \hat{\Phi} = 2 \kappa_{10}^2 g_s^2\, \Big[\,e^{2B} \Big( T(\Phi) +\frac{\partial T(\Phi)}{\partial \Phi}\Big) - \frac{\partial \mu(\Phi)}{\partial \Phi} \, \xi(r) - \frac{\partial \delta \mu_{\rm T}^m (\Phi)}{\partial \Phi}\,f_m (y) \,\Big] \,  \delta^6 (\vec{r})\,\,,
\end{equation}
and similarly for $\nabla^2 B$. Also the total brane action is
\begin{equation}
I_{\rm brane} =\big[\int d^4 x \, \sqrt{-g_4} \big]\,\int r^5 dr \epsilon_5 \,\Big[-e^{2B} T(\Phi) + \mu(\Phi) \xi(r) + \delta \mu_{\rm T}^m (\Phi) f_m (y) \Big]\,  \delta^6 (\vec{r}) \,\,.
\end{equation}

The last term of (6.7) and (6.8) will act as a supersymmetry breaking term (see Secs. 6.3 and 7.1). Since it occurs as a result of the gauge symmetry breaking of $A_4$, we may say that the primary cause of the supersymmetry breaking is a five-form anomaly in a sense. But this is not the usual gravitational anomaly for the five-form. In the usual gravitational anomaly the five-form acts as a source for the gravity. But in here the five-form plays the role of the gauge field coupled to the string fields on the D3-brane which now act as a source for the five-form. The gauge symmetry breaking of $A_4$ does not occur in the bulk region, so the bulk region does not suffer from this anomaly. From all this the anomaly associated with the supersymmetry breaking may have to be understood as a composition of the various anomalies for the string fields on the D3-brane which couple to $A_4$. The integration of these anomalies over the brane region turns out to vanish by the condition $\lambda=0$ (see Sec. IX). So the above result does not mean that the theory with broken supersymmetry becomes anomalous. In general the ten-dimensional superstring theory is known to be anomaly free \cite{21,22}. We will get back to this in Sec. IX.

\vskip 0.5cm
\hspace{-0.65cm}{\bf \large 6.2 D3-brane with a nonzero thickness}
\vskip 0.1cm

So far we have assumed that the D3-brane located at the conifold singularity has a zero thickness. But in reality D3-brane has its own thickness and the case III may not be suitable for a real model. For instance in (4.27) $M_{pl}$ diverges in the thin brane limit because $e^{-2B} \rightarrow \infty$ as $r \rightarrow 0$ in the case III of the previous section. By the same reason $\delta_Q I_{\rm brane}^{\rm (NS)}$ following from (4.11) with $T(\Phi) = T_0 \, e^{-\Phi} + \rho_{\rm vac}$ simply vanishes in the thin brane limit. This is not realistic because $\delta_Q I_{\rm brane}^{\rm (NS)}$ per unit volume of the four-dimensional spacetime must be at least of an order $\sim (TeV)^4$ as mentioned in Sec. IV. These unrealistic situations can be avoided if we allow for nonzero thickness to the brane because in that case the integration region does not contain the (neighborhood of the) point $\vec{r}=0$.

More importantly the supersymmetry breaking with vanishing $\lambda$ can be easily understood if we allow for nonzero thickness to the brane, as we shall see in this section a little later on. The scenario of nonzero thickness of D-brane associated with the cosmological constant problem has already been studied in the literature \cite{15,23}, where they argued that a tiny cosmological constant can be achieved if one assumes that the cosmological constant receives contributions only from the vacuum energy of the bulk fields. In the followings we will also assume that the D3-brane at the conifold singularity has a nonzero thickness $r_B$, though the entire structure of the mechanism is totally distinguished from that of \cite{15,23}. In this setup the delta function source terms of the field equations vanish outside the brane, while they do not inside the brane. In order to see it explicitly, we divide the whole transverse space into a brane region ($0 \leq r < r_B$) and bulk region ($r > r_B$), and modify the delta function $\delta^6 (\vec{r}\,)$ into
\begin{equation}
\delta^6 (\vec{r}\,) =
\begin{cases} \delta_0 & \text{for ~~$0 \leq r < r_B$} \\ 0 & \text {for~~~~~~~\,\, $r > r_B$\,\,,}
\end{cases}
~~~~~~~~\Bigg( \delta_0 \equiv \frac{6}{r_B^6} \, \frac{1}{\rm Vol(B)} \Bigg)\,\,.
\end{equation}

\vskip 0.5cm
\hspace{-0.65cm}{\bf \large 6.3 Field equations and their solutions in the brane region}
\vskip 0.1cm

In the brane region the whole field equations with delta function sources must be revised by (6.9). First, the field equation for $\xi (r)$ must be recast into
\begin{equation}
\frac{1}{r^5}\frac{d}{dr} \Big[e^{2\hat{\Phi}-4B}\,r^5 \frac{d\xi}{dr}\,\Big]=2 \kappa_{10}^2 \,\delta_0\, \mu(\Phi) \,\,,
\end{equation}
which, upon integration $\int r^5 dr \epsilon_5$, gives
\begin{equation}
\partial_r \xi = \frac{1}{g_s} \, \frac{{\rm q_{in}(r)}}{r^5}\,\chi \,\,,~~~~~~\Big({\rm q_{in}}(r) \equiv c_0 \int r^5 dr \mu (\Phi) \Big) \,\,,
\end{equation}
where $\chi= e^{4B- 2 \hat{\Phi}}$ as before and $c_0 = 2 \kappa_{10}^2 g_s \delta_0$. Using (6.11) one obtains from (6.7) and the remaining equations in (4.12)-(4.15):
\begin{equation}
\nabla^2 \hat{\Phi} = c_0 \,\chi^{1/2} e^{\Phi} \Big( \rho_{\rm vac} + \frac{\partial \rho_{\rm vac}}{\partial \Phi} \Big) -c_0\,g_s \Big[ \frac{\partial \delta \mu}{\partial \Phi} \xi + \frac{\partial \delta \mu_{\rm T}^m}{\partial \Phi} f_m \Big] \,\,,
\end{equation}
\begin{equation}
\nabla^2 \ln \chi - 2 \frac{[{\rm q}_{\rm in}(r)]^2} {r^{10}}\, \chi = 2 c_0\,\chi^{1/2} T_0 + 2 c_0\,\chi^{1/2} \,e^{\Phi}\rho_{\rm vac}  \,\,,
\end{equation}
\begin{equation}
\Big(\frac{d}{dr} \ln \chi \Big)^2 - 4 \frac{[{\rm q}_{\rm in}(r)]^2} {r^{10}}\, \chi + 4 \Big(\frac{d{\hat{\Phi}}}{dr}\Big)^2 =0 \,\,,
\end{equation}
where the condition $\beta =0$ with $R(r)=r$ is always understood in the above and in what follows.

The above equations are highly nonlinear in $e^{\Phi}$ and may be solved perturbatively (order by order) in $g_s$. Now we expand $e^{\hat{\Phi}}$ and $\chi^{1/2}$ as
\begin{equation}
e^{\hat{\Phi}}= e^{\hat{\Phi}_{(0)}} \Big(1+ g_s U_{(1)} + g_s^2 U_{(2)} + \cdots \Big)\,\,,
\end{equation}
\begin{equation}
\chi^{1/2} = \chi^{1/2}_{(0)} \Big(1+ g_s V_{(1)} + g_s^2 V_{(2)} + \cdots \Big)\,\,,
\end{equation}
and from $\mu (\Phi) = \sum_{n=0}^{\infty} \mu_n e^{n \Phi}$ we can write ${\rm q}_{\rm in}(r)$ as
\begin{equation}
{\rm q}_{\rm in}(r) = \frac{1}{6} c_0 \mu_0 r^6  +  g_s c_0 \mu_1 \int r^5 dr e^{\hat{\Phi}_{(0)}}+ \cdots  \,\,.
\end{equation}
Finally from (6.11) and (6.17)
\begin{equation}
\partial_r \xi = \frac{c_0 \mu_0}{6 g_s} r \chi_{(0)} + \frac{c_0}{6} r \chi_{(0)} \Big( \mu_1 + 2 \mu_0 V_{(1)} \Big) + \cdots  \,\,.
\end{equation}

\vskip 0.1cm
\hspace{-0.6cm}{\bf (A) Tree level}
\vskip 0.0cm

At the tree level of the D3-brane the vacuum energy terms $\rho_{\rm vac}$, $\delta \mu (\Phi)$, and $\delta \mu_{\rm T}^m (\Phi)$ all vanish and $T(\Phi)$ and $\mu({\Phi})$ are simply given by $T(\Phi) = T_0 e^{-\Phi}$ and $\mu (\Phi) =\mu_0$. The tree level equations are the lowest-order (order of $g_s^0$) equations of (6.12)-(6.14). They are
\begin{equation}
\nabla^2 \hat{\Phi}_{(0)} = 0 \,\,,
\end{equation}
\begin{equation}
\nabla^2 \ln \chi_{(0)} - 2 \, \frac{\rm q_0^2}{r_B^{12}} \, r^2 \chi_{(0)} = 2 c_0 T_0 \chi_{(0)}^{1/2} \,\,,
\end{equation}
\begin{equation}
\Big(\frac{d}{dr} \ln \chi_{(0)} \Big)^2 - 4 \, \frac{\rm q_0^2}{r_B^{12}} \, r^2 \chi_{(0)} + 4 \Big(\frac{d{\hat{\Phi}}_{(0)}}{dr}\Big)^2 =0 \,\,,
\end{equation}
where ${\rm q}_0 \equiv \frac{1}{6} c_0 r_B^6 \mu_0$.

The first equation is trivially solved by $\hat{\Phi}_{(0)} =0$, but the second equation has a nontrivial solution
\begin{equation}
\chi_{(0)}^{1/2} =\frac{c_{\rm H}}{r^2}\,\,,~~~~~~\Bigg(c_{\rm H} = \frac{18}{c_0 \mu_0} \Big(\frac{T_0}{\mu_0} \Big) \, \Bigg[ -1 \pm \sqrt{1- \frac{8}{9} \Big(\frac{\mu_0}{T_0} \Big)^2 } \,\Bigg]  \,\Bigg) \,\,
\end{equation}
for $\mu_0 \neq 0$. Substituting (6.22) into (6.21) shows that $\mu_0$ must be related to $T_0$ by $\mu_0 = T_0$. This accords with the result of the case II of Sec. V where the field equations require that the D3-brane should be a BPS state. Indeed one can check that (6.20) and (6.21) do not allow for non-BPS solution with $\mu_0 =0$ for the given value of $\hat{\Phi}=0$ as it should be. Thus $\chi_{(0)}^{1/2}$ is finally given by\footnote{The solution with minus sign in front of the square root in (6.22) does not satisfy (6.21).}
\begin{equation}
\chi_{(0)}^{1/2} = - \frac{12}{c_0 \mu_0}\, \frac{1}{r^2}\,\,, ~~~~~~(\mu_0 = T_0)\,\,,
\end{equation}
and from (6.18) $\xi(r)$ is now written as
\begin{equation}
\xi = \frac{1}{g_s^2} \Big[ \xi_{(0)} + g_s \xi_{(1)} + \cdots \Big]
\end{equation}
where $\xi_{(0)}$ and $\xi_{(1)}$ are defined, respectively, by
\begin{equation}
\xi_{(0)} =  - \frac{12}{c_0 \mu_0}\, \frac{ g_s}{r^2}\,\,,~~~~~~\xi_{(1)} =  - \frac{12 g_s}{c_0 \mu_0} \Bigg[ \Big(\frac{\mu_1}{\mu_0}\Big)\frac{1}{r^2} - 4 \int \frac{dr}{r^3} \, V_{(1)} \Bigg]\,\,.
\end{equation}

Brane action also can be expressed in a series form. From (6.8) and the given expansions of the previous functions, one can write $I_{\rm brane}$ as
\begin{equation}
I_{\rm brane} = \frac{1}{g_s^2} \Big[ I_{\rm brane}^{(0)} + g_s  I_{\rm brane}^{(1)} + \cdots \Big]\,\,,
\end{equation}
where
\begin{equation}
I_{\rm brane}^{(0)} = \delta_0 \, \big[\int d^4 x \, \sqrt{-g_4}\, \big]\,\int r^5 dr \epsilon_5 \Big[-g_s  T_0 \chi_{(0)}^{1/2} + \mu_0 \xi_{(0)} \Big]
\end{equation}
is the tree level action of the order $g_s^0$, and
\begin{equation}
I_{\rm brane}^{(1)} = \delta_0 \, \big[\int d^4 x \, \sqrt{-g_4} \, \big]\,\int r^5 dr \epsilon_5 \Big[-g_s \chi_{(0)}^{1/2} (V_{(1)} T_0 +\rho_0 )+ \mu_0 \xi_{(1)} + \mu_1 \xi_{(0)} + \nu_1^m f_m^{(0)}  \Big]
\end{equation}
is the one-loop action of the order $g_s^1$, where the integration $\int r^5 dr \epsilon_5$ is taken over the brane region and we have set
\begin{equation}
\delta \mu_{\rm T}^m (\Phi) = \sum_{n=1}^{\infty} \nu_n^m e^{n \Phi}\,\,,~~~~~~f_m =  \frac{1}{g_s^2} \Big[ f_m^{(0)} + g_s f_m^{(1)} + \cdots \Big]\,\,,
\end{equation}
as in $\delta \mu (\Phi)$ and $\xi(r)$. The condition $\lambda =0$ requires that all $I_{\rm brane}^{(n)}$'s in (6.26) should vanish respectively. At the tree level, $I_{\rm brane}^{(0)}$ vanishes if $\mu_0 =T_0$ because $\chi_{(0)}^{1/2}$ and $\xi_{(0)}$ satisfy the relation $g_s \chi_{(0)}^{1/2} = \xi_{(0)}$ by (6.23) and the first equation of (6.25). Thus the condition $\lambda =0$ at the tree level requires that the D3-brane should be a BPS state as in the previous section.

\vskip 0.1cm
\hspace{-0.6cm}{\bf (B) One-loop level}
\vskip 0.0cm

One-loop level equations can be obtained by collecting $g_s^1$-order terms in (6.12)-(6.14). They are
\begin{equation}
\nabla^2 U_{(1)} = \frac{c_0}{g_s} \Big( \rho_0 g_s \chi_{(0)}^{1/2} - \mu_1 {\xi_{(0)}} \Big) - \frac{c_0}{g_s} \nu_1^m f_m^{(0)} \,\,,
\end{equation}
\begin{equation}
\nabla^2 V_{(1)} - \frac{c_0^2 \mu_0^2}{18}\, \Big( \frac{\mu_1}{\mu_0} + V_{(1)} \Big) r^2 \chi_{(0)}= c_0 \big(T_0 V_{(1)} + \rho_0 \big) \chi_{(0)}^{1/2} \,\,,
\end{equation}
\begin{equation}
\Big(\frac{d}{dr} \ln \chi_{(0)} \Big) \Big( \frac{dV_{(1)}}{dr} \Big) - \frac{c_0^2 \mu_0^2}{18} \, \Big( \frac{\mu_1}{\mu_0} + V_{(1)} \Big) r^2 \chi_{(0)} =0 \,\,.
\end{equation}
The first two terms on the right-hand side of (6.30) cancel by (6.23) and (6.25) for $\rho_0 =\mu_1$, and (6.30) reduces to
\begin{equation}
\nabla^2 {\hat \Phi} =  - {c_0} \, \rho_T^{(1)} \,\,,
\end{equation}
where we have used the fact that $\hat{\Phi} = g_s U_{(1)}$ in the first-order (one-loop level) approximation, and $\rho_T^{(1)}$ is defined by
\begin{equation}
\rho_T^{(1)} = \nu_1^m f_m^{(0)} \,\,.
\end{equation}
$\rho_T^{(1)}$ is one of the (one-loop order) vacuum energy density whose distribution is given by $f_m^{(0)}$. It originates from the quantum excitations with components along the transverse directions and becomes the primary cause of the supersymmetry breaking. Were it not for this term, the supersymmetry would be unbroken for $\rho_0 = \mu_1$, which is the $n=0$ case of (5.26). But since $\rho_T^{(1)} \neq 0$ in general, the cosmological constant problem requires that the condition $\lambda =0$ must be compatible with the nonzeroness of $\rho_T^{(1)}$. We will be back to this in Sec. 7.1.

The solution to the remaining equations (6.31) and (6.32) is found to be
\begin{equation}
V_{(1)} = -\frac{\mu_1}{\mu_0} \Big( 1 + \frac{r_0^2}{r^2} \Big) \,\,,
\end{equation}
where $r_0$ is an arbitrary constant and the condition $\rho_0 = \mu_1$ is also required by (6.32). The solution (6.35) is valid for any $U_{(1)}$ because (6.31) and (6.32) do not depend on $U_{(1)}$. This then implies that both supersymmetry-broken and unbroken solutions described by the same ($\chi_{(0)}$, $V_{(1)}$, $\xi_{(0)}$, $\xi_{(1)}$) are equally qualified for a solution to the field equations. This a little unexpected result is due to the fact that in the $g_s^1$-order approximation the $U_{(1)}$-terms in $e^{\Phi}$ and $(\partial \hat{\Phi})^2$ of (6.13) and (6.14) are already of an order $g_s^2$ and do not appear in the $g_s^1$-order equations of (6.13) and (6.14). The same thing also happens in the higher-order equations. In the $g_s^n$-order equations $U_{(n)}$ appears as of an order $g_s^{n+1}$ in $e^{\Phi}$ and similarly as of an order $g_s^{2n}$ in $(\partial \hat{\Phi})^2$, respectively. So $U_{(n)}$ does not appear in the $n$th order equations of (6.13) and (6.14), and these equations admit both solutions (the one with $U_{(n)} =0$ and the other with $U_{(n)} \neq 0$) to a correct solution.

\vskip 1cm
\hspace{-0.65cm}{\bf \Large {VII. Address the cosmological constant problem}}
\vskip 0.5cm
\setcounter{equation}{0}
\renewcommand{\theequation}{7.\arabic{equation}}

In the previous section we have considered a configuration of D3-brane which possesses its own thickness, and then found solutions to the field equations in the brane region. So in the next we may have to work out the field equations in the bulk region. However, we do not need to do this. The field equations and their solutions in the bulk region are already found in the case III of the Sec. V. The supersymmetry is unbroken and $\lambda =0$ is maintained by field equations. So we do not need to know about the solutions of the bulk region anymore. The solutions of the brane region obtained in Sec. 6.3 is sufficient enough to address the cosmological constant problem.

\vskip 0.5cm
\hspace{-0.65cm}{\bf \large 7.1 Supersymmetry breaking with vanishing ${\lambda}$}
\vskip 0.2cm

In Sec. 6.3 we have seen that the tree level action $I_{\rm brane}^{(0)}$ vanishes for $\mu_0 =T_0$, and the field equation for ${\hat \Phi}$ possesses a supersymmetry breaking term at the one-loop level (see (6.33)). Now $\lambda =0$ requires\footnote{Recall that $\lambda$ must vanish by the condition $\beta =0$ which is required by (4.20).} that $I_{\rm brane}^{(1)}$ must also vanish. To see this, we substitute (6.35) into the second equation of (6.25) to get
\begin{equation}
\xi_{(1)} = \frac{12 g_s}{c_0 \mu_0} \Big( \frac{\mu_1}{\mu_0} \Big) \frac{1}{r^2} \Big( 1+ \frac{r_0^2}{r^2} \Big) = g_s \chi_{(0)}^{1/2}V_{(1)}\,\,.
\end{equation}
Using (7.1) and $g_s \chi_{(0)}^{1/2} = \xi_{(0)}$ one finds that all but the last term in (6.28) cancel out for $\mu_0 =T_0$ and $\mu_1 =\rho_0$, and we are left with
\begin{equation}
I_{\rm brane}^{(1)} = \delta_0 \, \big[\int d^4 x \, \sqrt{-g_4} \, \big]\,\int r^5 dr \epsilon_5 \,\nu_1^m f_m^{(0)} \,\,.
\end{equation}
Thus when we consider up to the one-loop level, $\lambda$ can be written as
\begin{equation}
\lambda =\frac{\kappa^2}{2} \delta_0 \int r^5 dr \epsilon_5 \, \rho_T^{(1)} \,\,,
\end{equation}
where the integration is taken over the whole brane region of the transverse space.

Now the point of the cosmological constant problem can be summarized as whether we can find a nonzero function $\rho_T^{(1)}$ satisfying
\begin{equation}
\delta_0 \int_{r=0}^{r=r_B} r^5 dr \epsilon_5 \, \rho_T^{(1)} \equiv Q_{\rm total}^{T}= 0 \,\,,
\end{equation}
where $Q_{\rm total}^{T}$ represents the total vacuum energy (per unit volume of the four-dimensional spacetime) of the brane region which originated from the excitation $<\chi_{\rm vac}^{m123}>$. For such a function $\rho_T^{(1)}$, $\lambda$ vanishes by (7.4) because $\lambda$ is given by $\lambda = (\kappa^2 /2) \, Q_{\rm total}^{T}$, and the supersymmetry is broken by (6.33) because $\rho_T^{(1)}$ is a nonzero function, which is the very configuration required by the cosmological constant problem. In the followings we will consider two different cases of this configuration.

First we integrate (6.33) by $\int d^6 y \sqrt{h_6}$ and use the divergent theorem to find $\partial_m {\hat \Phi}$, where $h_{mn}$ represents the conifold metric (2.2). We obtain
\begin{equation}
\oint_{S(y^m)} d \Sigma_m \sqrt{h_6} \,\, h^{mn} \partial_n {\hat \Phi} = - c_0 \int_{0}^{y^m} d^6 y \sqrt{h_6} \, \rho_T^{(1)} \,\,,
\end{equation}
where $\oint_{S(y^m)} d \Sigma_m$ is a surface integral taken over the hypersurface defined by a certain fixed $y^m$. Now let us find the configurations which respect (7.4) as required by the cosmological constant problem.

\vskip 0.1cm
\hspace{-0.6cm}{\bf (A) Case I}
\vskip 0.0cm

Suppose that ${\hat \Phi}$ is only a function of $r$: ${\hat \Phi} = {\hat \Phi}(r)$. In this case (7.5) reduces to
\begin{equation}
\partial_r {\hat \Phi} = - \frac{c_0}{r^5} \int_{0}^{r} r^5 dr \rho_T^{(1)} \,\,,
\end{equation}
and the natural solution $\rho_T^{(1)}$ satisfying (7.4) may be written as
\begin{equation}
\rho_T^{(1)} = \frac{r_B}{r^5} \sum_{n=1}^{\infty}  (a_n \cos k_n r + b_n \sin k_n r)\,\,,~~~~~~ \Big(k_n \equiv \frac{2 \pi n}{r_B} \Big) \,\,,
\end{equation}
where $a_n$ and $b_n$ are arbitrary dimensionless constants. Substituting (7.7) into (7.6) one obtains
\begin{equation}
\partial_r {\hat \Phi} = {c_0}\, \frac{r_B}{r^5} \sum_{n=1}^{\infty} \Big[ \frac{1}{k_n} (b_n \cos k_n r - a_n \sin k_n r) - \frac{b_n}{k_n} \Big]\,\,.
\end{equation}
Note that $\partial_r \hat{\Phi}$ without the last (constant) term of (7.8) is also a good solution to (6.33) corresponding to (7.7). This solution is interesting because it diverges as $r \rightarrow 0$ meaning that the magnitude of the supersymmetry breaking is infinitely large at $r=0$.

\vskip 0.1cm
\hspace{-0.6cm}{\bf (B) Case II}
\vskip 0.0cm

Now suppose that $<\chi_{\rm vac}^{m123}>$ contains the excitations along the isometry directions $\psi$ or $\phi_i$: $m= \psi$ or $\phi_i$. In this case $\rho_T^{(1)}$ becomes a function of $\psi$ or $\phi_i$ (and necessarily of $r$ and $\theta_i$). Choose $m= \psi$ for definiteness of our discussion and suppose that ${\hat \Phi}$ is a function of $\psi$ alone\footnote{In this case the $\psi$ dependence of $\hat \Phi$ must be taken into account when we solve the field equations (for $\chi$) with order higher than $g_s^2$. But the field equations in Sec. 6.3 remain unchanged because they are $g_s^0$- and $g_s^1$-order equations.}: ${\hat \Phi} = {\hat \Phi}(\psi)$. For this ${\hat \Phi}$ (7.5) reduces to
\begin{equation}
h^{\psi \psi} \partial_{\psi} {\hat \Phi} = - c_0 \int_{0}^{\psi} d \psi \, \rho_T^{(1)} \,\,,
\end{equation}
and $\rho_T^{(1)}$ satisfying (7.4) is now given by
\begin{equation}
\rho_T^{(1)} =  f(r, \theta_{i}) \sum_{n=1}^{\infty} \Big[ a_n \cos \frac{n}{2} \psi + b_n \sin \frac{n}{2} \psi \Big]\,\,,~~~~~~(-2 \pi \leq \psi \leq 2 \pi ) \,\,,
\end{equation}
where $a_n$ and $b_n$ are arbitrary dimensionless constants, while $f(r, \theta_{i})$ is an arbitrary function of $r$ and $\theta_i$ with length dimension minus four. Also $f(r, \theta_i )$ must satisfy the condition that $r^5 f(r, \theta_{i})$ be regular in the region $0 \leq r \leq r_B$.

For $f(r, \theta_{i})= - h^{\psi \psi} / {r_0^2 }$ (where $r_0$ is an arbitrary constant with length dimension one) (7.9) is solved by
\begin{equation}
\partial_{\psi} {\hat \Phi} = \gamma_0 \sum_{n=1}^{\infty} \Big[ \tilde{a}_n \cos \frac{n}{2} \psi + \tilde{b}_n \sin \frac{n}{2} \psi \Big]\,\,,~~~~~\Big(\gamma_0 \equiv  \frac{c_0}{r_0^2} \Big)\,\,,
\end{equation}
and therefore
\begin{equation}
{\hat \Phi} = -\gamma_0 \sum_{n=1}^{\infty} \Big( \frac{2}{n}\Big)^2 \Big[ {a}_n \cos \frac{n}{2} \psi + {b}_n \sin \frac{n}{2} \psi \Big]\,\,,
\end{equation}
where $\tilde{a}_n \equiv -(2/n) b_n$, $\tilde{b}_n \equiv (2/n) a_n$, and $\gamma_0$ is a dimensionless constant of order $g_s^1$. (7.11) still represents an arbitrary Fourier expansion because $\tilde{a}_n$ and $\tilde{b}_n$ are arbitrary as well as $a_n$ and $b_n$.

\vskip 0.1cm
\hspace{-0.6cm}{\bf (C) Orbifold compactifications}
\vskip 0.0cm

There is a simple but important solution subject to (7.11). The step function $\Theta(\psi)$ defined by $\Theta(\psi) = +1(-1)$ for $0 < \psi < 2 \pi$ ($- 2 \pi < \psi < 0$) can be represented by the sine series ($\tilde{a}_n =0$) of (7.11). In this representation ${\hat \Phi}$ is given by ${\hat \Phi} = \gamma_0 |\psi|$ and therefore $e^{\Phi} = g_s e^{\gamma_0 |\psi|}$. Now rewrite (4.3) (with (4.9)) as
\begin{equation}
ds_{10}^2 = e^{-B} d \tilde{s}_{\rm conifold}^2 + e^{B} g_{\mu\nu}dx^{\mu} dx^{\nu} \,\,,
\end{equation}
where $d \tilde{s}_{\rm conifold}^2$ is defined by $d \tilde{s}_{\rm conifold}^2 = e^{\hat{\Phi}} ds_{\rm conifold}^2$. Then taking $\gamma_0 =1$ (namely take $r_0^2 =c_0$) one obtains
\begin{equation}
d \tilde{s}_{\rm conifold}^2  =  e^{|\psi|} \, ds_{\rm conifold}^2 \,\,.
\end{equation}
(7.14) represents a $\mathbb{Z}_2$-orbifolding where the conifold metric is $\mathbb{Z}_2$-orbifolded along $\psi$ with $\psi \cong -\psi$. In the same way if we introduce a new step function $\Theta(\psi) = +1(-1)$ for $0 < \psi < \frac{2 \pi}{N}$ ($- \frac{2 \pi}{N} < \psi < 0$) mod $4 \pi$ we can generalize (7.14) to the $\mathbb{Z}_N$-orbifolding generated by ($r,t^m$) where $r$ is the reflection $\psi \cong -\psi$ and $t^m$ is the translation $\psi \cong \psi + \frac{4 \pi}{N}m$. We can do the same thing to the case of $m=\phi_i$ to orbifold the metric along $\phi_i$.

Orbifold compactification generally reduces the number of unbroken supersymmetries of the theory and in our case one can expect that the $\mathbb{Z}_N$-orbifolding leads to a nonsupersymmetric theory because originally we had only one (for instance ${\mathcal N}=1$, $d=4$ for the type IIB) unbroken supersymmetry which however would be broken by the $\mathbb{Z}_N$-orbifolding. Good examples are given in \cite{24} where the authors presented a nonsupersymmetric type II theory compactified on orbifolded $T_6$ in which the supersymmetry is broken at the string scale but the quantum corrections to the cosmological constant cancel. These examples may explain why $ds_{\rm conifold}^2$ in $ds_{10}^2$ should be replaced by $d \tilde{s}_{\rm conifold}^2$ in (7.13). However, the usual orbifolding methods \cite{25} including \cite{24} are no more than a special way to obtain nonsupersymmetric theories with vanishing $\lambda$. According to the discussion of Sec. 7.1 the most general way to obtain nonsupersymmetric theory with vanishing $\lambda$ is to replace $ds_{\rm internal}^2$ $\rightarrow$ $e^{\hat \Phi} ds_{\rm internal}^2$, where ${\hat \Phi}$ is a solution to (6.33) and where $\rho_T^{(1)}$ is subject to (7.4).

The whole discussion in Sec. 7.1 is basically based on the assumption that the D3-brane has a nonzero thickness $r_B$. Indeed in the case I, $k_n$ diverges as $r_B \rightarrow 0$ and hence ${\hat \Phi}$ or its derivative $\partial {\hat \Phi}$ is not well defined in the thin-brane ($r_B =0$) limit. However, ${\hat \Phi}$ and the other functions of the case II do not contain the parameter $r_B$ and therefore taking the limit $r_B \rightarrow 0$ certainly makes sense in the case II. So the whole discussions of this paper may not be restricted only to the case of the brane with nonzero thickness.

\vskip 0.5cm
\hspace{-0.65cm}{\bf \large 7.2 Why broken supersymmetry?}
\vskip 0.2cm

In Sec. 7.1 we have seen that the $d=4$ reduced supersymmetry can be arbitrarily broken (while maintaining $\lambda=0$) if $\rho_T^{(1)}$ is given by the Fourier expansions satisfying (7.4). However, the Fourier expansions contain a special case, $\rho_T^{(1)} =0$, in which the coefficients $a_n$ and $b_n$ all vanish. In this case $\lambda$ of course vanishes by (7.3), but at the same time the $d=4$ supersymmetry also remains unbroken. This is certainly not the case of our universe because the $d=4$ supersymmetry of our universe is believed to be broken. So the question is that why our universe did not make this relatively simple choice. Why did our universe choose a nonzero $f_m^{(0)}$? We may have to answer this last question to make our discussion on the cosmological constant problem complete.

To find the answer to this question, we go back to (4.10) and (6.8) to calculate the minimum value of the total action $I_{\rm total}$. The brane action (6.8) vanishes in any case by the condition $\lambda =0$ or (7.4). Also the first term of (4.10) vanishes for $\lambda =0$ since ${\mathcal R}_4 (g_{\mu\nu})$ is proportional to $\lambda$. Thus the minimum value of $I_{\rm total}$ is entirely determined by the remaining terms of (4.10), which is just given by (4.26). Since ${\mathcal R}_6 (h_{mn})$ vanishes for $R(r)=r$ the total action finally becomes
\begin{equation}
I_{\rm total} \propto \int d^6 y \sqrt{h_6} \Big[ - \frac{1}{8} \big( \partial \ln \chi)^2 + \frac{g_s^2}{2} \chi^{-1} (\partial \xi)^2   - \frac{1}{2} \big( \partial \hat{\Phi} \big)^2 \,\Big] \,\,,
\end{equation}
where the positive proportionality constant $\big[ \int d^4 x \sqrt{-g_4} \,\big]/ 2 \kappa_{10}^2 \,g_s^2$ has been omitted. Basically (7.15) vanishes by (6.14) to all orders of $g_s$. However, (6.14) is too restrictive in the phenomenological sense because it was obtained from the conventional action with the smallest number of field contents. In more realistic extended models we need more terms in the action. So if we can relax (6.14) at the order of (and higher than) $g_s^2 \,$ the following discussion can be made.

(7.15) can be expanded in a power series of $g_s$ as
\begin{equation}
I_{\rm total} \propto \int d^6 y \sqrt{h_6} \Big[ \Sigma_{(0)} + {g_s} \Sigma_{(1)} - \frac{g_s^2}{2} \big( \partial U_{(1)} \big)^2 + \cdots \,\Big] \,\,,
\end{equation}
where
\begin{equation}
\Sigma_{(0)} = - \frac{1}{8} \big( \partial \ln \chi_{(0)} \big)^2 + \frac{1}{2 g_s^2}\,  \big( \partial \xi_{(0)} \big)^2  \chi_{(0)}^{-1}\,\,,
\end{equation}
\begin{equation}
\Sigma_{(1)} = - \frac{1}{2} \big( \partial \ln \chi_{(0)} \big) \big( \partial V_{(1)} \big) + \frac{1}{g_s^2} \, \chi_{(0)}^{-1} \Big[ \big( \partial \xi_{(0)} \big)\big( \partial \xi_{(1)} \big)- V_{(1)} \big( \partial \xi_{(0)} \big)^2  \Big] \,\,.
\end{equation}
In Sec. 6.3 we have seen that the nonsupersymmetric solution described by ($\chi_{(0)}$, $V_{(1)}$, $\xi_{(0)}$, $\xi_{(1)}$) with nontrivial $U_{(1)}$ is as equally qualified for a solution to the field equations as the supersymmetric solution described by the same ($\chi_{(0)}$, $V_{(1)}$, $\xi_{(0)}$, $\xi_{(1)}$) but with vanishing $U_{(1)}$. Both of these solutions can satisfy the field equations because $\chi_{(0)}$ and $V_{(1)}$ (and consequently $\xi_{(0)}$ and $\xi_{(1)}$) are entirely determined by the $U(1)$-independent field equations, and on the other hand $U_{(1)}$ is only determined by an arbitrary $f_m^{(0)}$ through (6.30). Now one can check that both $\Sigma_{(0)}$ and $\Sigma_{(1)}$ vanish for the given ($\chi_{(0)}$, $V_{(1)}$, $\xi_{(0)}$, $\xi_{(1)}$) of Sec. 6.3. So the first two terms in (7.16) vanish for both solutions, and $I_{\rm total}$ is finally given by
\begin{equation}
I_{\rm total} \propto - \frac{g_s^2}{2} \int d^6 y \sqrt{h_6} \,(\partial U_{(1)})^2 + \cdots \,\,.
\end{equation}
The first term of (7.19) takes negative values for the solutions with nontrivial $U_{(1)}$, while it vanishes for the solution with vanishing $U_{(1)}$. Thus (7.19) suggests that the solution with broken supersymmetry would be more favored by the action principle than the other with unbroken supersymmetry because the former takes lower values of $I_{\rm total}$ than the latter. But see also the fourth last paragraph of Sec. IX.

\vskip 1cm
\hspace{-0.65cm}{\bf \Large {VIII. With nonvanishing fluxes}}
\vskip 0.5cm
\setcounter{equation}{0}
\renewcommand{\theequation}{8.\arabic{equation}}

So far in the previous sections we have considered the conventional compactifications with $H_3 =0$ and $\Phi={\rm constant}$ to address the cosmological constant problem. In the framework of these compactifications we found that $\lambda$ adjusts itself to zero forced by field equations, while the $d=4$ supersymmetry is broken in the brane region. But the conventional compacitifications typically suffer from having too many moduli whose vacuum expectation values are not properly determined. To obtain more realistic phenomenological models we may need to generalize our discussion to the case where we have some nontrivial potential that can freeze these undetermined moduli. In this section we will consider the flux compactifications where the fluxes including $H_3$ are all turned on to stabilize the moduli of the Calabi-Yau threefolds.

\vskip 0.5cm
\hspace{-0.65cm}{\bf \large 8.1 Superpotential and scalar potential}
\vskip 0.2cm

Consider the low-energy effective action of the type IIB string theory. In the Einstein frame it can be written in the $SL(2, {\mathbb R})$ invariant form,
\begin{eqnarray}
I_{\rm IIB}= \frac{1}{2 \kappa_{10}^2} \int d^{10} x \sqrt{-G} \Big[ \mathcal{R}_{10} - \frac{(\nabla \phi)^2}{2(\rm Im \phi)^2} - \frac{1}{2 \cdot 3!} \frac{G_3 \cdot \bar{G}_3}{\rm Im \phi} - \frac{1}{2 \cdot 5!} {\tilde F}_5^2 \Big] \nonumber \\ \nonumber \\
+ \frac{2}{2 \kappa_{10}^2} \int \frac{A_4 \wedge G_3 \wedge {\bar G}_3}{4 i \rm Im \phi} ~~~~~~~~~~~~~~~~~~~~~~~~~~~~~~~~~~~~~~~
\end{eqnarray}
with
\begin{equation}
G_3 = F_3 - \phi H_3\,\,, ~~~~~~~~\phi= A_0 + i e^{-\Phi}\,\,,
\end{equation}
and
\begin{equation}
\tilde{F} = F_5 - \frac{1}{2} A_2 \wedge H_3 + \frac{1}{2} B_2 \wedge F_3 \,\,,~~~~~~~~ {}^{*}{\tilde F}_5 = {\tilde F}_5 \,\,,
\end{equation}
where $F_3$($=dA_2$) is the R-R three-form field strength, and $\phi$ is the axion/dilaton. The $G_3 \cdot \bar{G}_3$ term in (8.1) gives rise to a potential for $\phi$ and the complex structure moduli of the Calabi-Yau threefold as we shall see below. The last term, which is the Chern-Simons term, makes a contribution to the total D3 charge, but is irrelevant to the Einstein equation because it does not contain the metric.

In the presence of nonzero $G_3$, one can generate a superpotential $W$ for the Calabi-Yau moduli \cite{26} as
\begin{equation}
W=\int_{\mathcal{M}_6} G_3 \wedge \Omega \,\,,
\end{equation}
where $\mathcal{M}_6$ is the Calabi-Yau threefold and $\Omega$ is its holomorphic three-form. (8.4) shows that $W$ vanishes if $G_3$ does not contain $(0,3)$ component. In particular, for $(2,1)$ type $G_3$, $W$ satisfies (for instance see \cite{27}),
\begin{equation}
W= \partial_{\phi} W = \partial_{\tau_i} W = 0 \,\,,
\end{equation}
which corresponds to an unbroken supersymmetry and where $\tau_i$ are complex structure moduli of $\mathcal{M}_6$. Aside from this, the $G_3 \cdot \bar{G}_3$ term in (8.1) can be written as \cite{28}
\begin{equation}
I_{\rm G} = \int d^4 x \sqrt{-g_4} \, \mathcal{L}_{\rm G}
\end{equation}
with
\begin{eqnarray}
\mathcal{L}_{\rm G} = \frac{1}{4 \kappa_{10}^2} \int_{\mathcal{M}_6} \frac{G_3 \wedge *_6 \bar{G}_3}{\rm Im \phi}~~~~~~~~~~~~~~~\,  \nonumber\\\nonumber\\
= \mathcal{V}_{\rm scalar} - \frac{i}{4 \kappa_{10}^2 \rm Im \phi} \int_{\mathcal{M}_6} G_3 \wedge \bar{G}_3 \,\,,
\end{eqnarray}
where $*_6$ is the dual in the transverse directions, and $\mathcal{V}_{\rm scalar}$ is given by
\begin{equation}
\mathcal{V}_{\rm scalar} = \frac{1}{2 \kappa_{10}^2 \rm Im \phi} \int_{\mathcal{M}_6} G_3^{\rm IASD} \wedge *_6 \bar{G}_3^{\rm IASD} \,\,.
\end{equation}
In (8.8) $G_3^{\rm IASD}$ is the imaginary anti self-dual (IASD) part of $G_3$, $*_6 {G}^{\rm IASD} = -i G^{\rm IASD}$, and the second term of (8.7) is merely topological and does not involve any moduli.

Defining the K$\ddot{\rm a}$hler potential $\mathcal{K}$ as
\begin{equation}
\mathcal{K} = - \ln [-i (\phi - \bar{\phi})] - \ln \Big[ -i \int_{\mathcal{M}_6} \Omega \wedge {\bar \Omega}\, \Big] \,\,,
\end{equation}
one can show \cite{28} that $\mathcal{V}_{\rm scalar}$ can be expressed in terms of $W$ as
\begin{equation}
\mathcal{V}_{\rm scalar} = \frac{1}{2 \kappa_{10}^2}\, e^{\mathcal{K}} \,\Big[ {\mathcal G}^{i {\bar j}} D_i W \,\overline{D_j W} \,\Big] \,\,,
\end{equation}
where $D_i W =\partial _i W + (\partial_i \mathcal{K}) W$ and ${\mathcal G}_{i {\bar j}} = \partial_i \partial_{\,\bar j} \mathcal{K}$, and $i$, $j$ are summed over $\phi$ and the complex structure moduli $\tau_i$. (8.10), however, is not the most general expression for the potential. To obtain a general expression for the potential we need to introduce the K$\ddot{\rm a}$hler potential for the K$\ddot{\rm a}$hler moduli which generally runs over up to $h_{1,1}$. In our discussion we just assume that we have only one K$\ddot{\rm a}$hler modulus, say $\rho$. For such a model the tree level K$\ddot{\rm a}$hler potential takes the form
\begin{equation}
\mathcal{K} = -3 \ln [-i (\rho - \bar{\rho})] - \ln [-i (\phi - \bar{\phi})] - \ln\Big[ -i \int_{{\mathcal M}_6} \Omega \wedge {\bar \Omega}\, \Big] \,\,,
\end{equation}
and similarly (8.10) is generalized to the form
\begin{equation}
\mathcal{V}_{\rm scalar} = \frac{1}{2 \kappa_{10}^2} \,e^{\mathcal{K}} \, \Big[ {\mathcal G}^{a {\bar b}} D_a W \, \overline{D_b W} -3 |W|^2 \,\Big] \,\,,
\end{equation}
where $a$, $b$ are now summed over $\rho$ as well as $\phi$ and $\tau_i$.

\vskip 0.5cm
\hspace{-0.65cm}{\bf \large 8.2 $\bf \lambda$ in flux compactifications}
\vskip 0.2cm

The effective bulk action (8.1) can be rewritten in the string frame as
\begin{eqnarray}
I_{\rm IIB}= \frac{1}{2 \kappa_{10}^2} \int d^{10} x \sqrt{-G} \Bigg[ e^{-2\Phi} \Big[\mathcal{R}_{10} + 4 (\nabla \Phi)^2 \Big] - \frac{1}{2}F_1^2 - \frac{1}{2 \cdot 3!} {G_3 \cdot \bar{G}_3} - \frac{1}{2 \cdot 5!} {\tilde F}_5^2 \Bigg]
\nonumber \\ \nonumber \\
+ \frac{1}{8i \kappa_{10}^2} \int e^{\Phi} {A_4 \wedge G_3 \wedge {\bar G}_3}\,\,,~~~~~~~~~~~~~~~~~~~~~~~~~~~~~~~~~~~~~~~~~~~~~~~~~
\end{eqnarray}
where the metric $G_{MN}$ and the Ricci scalar $\mathcal{R}_{10}$ are now those of the string frame. (8.13) shows that in the presence of nonzero fluxes the six-dimensional bulk action will generally take the form
\begin{eqnarray}
I_{\rm bulk}/[\int d^{4}x \sqrt{-g_4}]=\frac{1}{2\kappa_{10}^2 g_s^2} \int d^{6}y \sqrt{h_6} \, e^{{\hat\Phi}-2B} \beta ~~~~~~~~~~~~~~~~~~~~~~~~~~~~~~~~~~\nonumber\\\nonumber\\
~~~~~~~~~~~~~~~+ \frac{1}{2\kappa_{10}^2 g_s^2} \int d^{6}y \sqrt{h_6}\, \big[{\mathcal R}_6 (h_{mn}) - L_F \big] + {\rm topological~terms}\,\,
\end{eqnarray}
in the string frame, where $\beta = [\int d^4 x \sqrt{-g_4} {\mathcal R}_4 ]/[\int d^4 x \sqrt{-g_4}]$ as before and
\begin{equation}
L_{F}= K-V \,\,, ~~~~~\big(K=h^{mn} K_{mn} \big)\,\,,
\end{equation}
with
\begin{equation}
K_{mn}= \sum_{I,J} F_{IJ} [\phi_K ] \, \partial_{(m} \phi_{{}_{I}} \partial_{n)} \phi_{{}_{J}} \,\,, ~~~V=V[\phi_{I}, h^{mn}] \,\,,
\end{equation}
where $\phi_I$'s are six-dimensional scalar fields including $\Phi$, $B$, $\xi$ etc. (8.14) is a generalization of (4.10), and in (8.15) $V$ is related to $\mathcal V_{\rm scalar}$ by the equation
\begin{equation}
\mathcal V_{\rm scalar} = \frac{1}{2\kappa_{10}^2 g_s^2} \int d^{6}y \sqrt{h_6}\,\, V \,\,,
\end{equation}
and hence $V$ is identified with $-\frac{g_S^2}{3!}\, G_3^{\rm IASD} \cdot {\bar G}_3^{\rm IASD}$ of the type IIB action (8.13). The six-dimensional Einstein equation following from (8.14) is now
\begin{equation}
\mathcal R_{mn} - \frac{1}{2} h_{mn} \mathcal R_6 - \frac{1}{2} T_{mn} - \frac{\beta}{2}  e^{{\hat\Phi}-2B} h_{mn}=0 \,\,,
\end{equation}
where the energy momentum tensor $T_{mn}$ is defined by
\begin{equation}
T_{mn} = \frac{2}{\sqrt{h_6}} \frac{\delta (\sqrt{h_6}\,L_F)}{\delta h^{mn}}\,\,.
\end{equation}

The four-dimensional Einstein equation on the other hand can be obtained by rewriting (8.14) as
\begin{equation}
I_{\rm bulk} = \frac{1}{2\kappa^2} \int d^4 x \sqrt{-g_4}\, {\mathcal R}_4 (g_{\mu\nu}) + \int d^4 x \sqrt{-g_4} \,{\hat I}_{\rm bulk} + {\rm topological~terms}\,\,,
\end{equation}
where ${\hat I}_{\rm bulk}$ is now
\begin{equation}
{\hat I}_{\rm bulk} = \frac{1}{2\kappa_{10}^2 g_s^2} \int d^{6} y \sqrt{h_6} \big[{\mathcal R}_6 (h_{mn}) - L_F \big] \,\,,
\end{equation}
which is the generalization of (4.26). Adding $I_{\rm brane}$ to (8.20) one finds that $I_{\rm total}$ takes the form
\begin{equation}
I_{\rm total} = \frac{1}{2\kappa^2} \int d^4 x \sqrt{-g_4} \big[{\mathcal R}_4 (g_{\mu\nu}) - 2 \lambda \big] + {\rm topological~terms}\,\,,
\end{equation}
where $\lambda$ is defined by
\begin{equation}
\lambda = - \kappa^2 \big[ {\hat I}_{\rm bulk} + {\hat I}_{\rm brane} \big]\,\,
\end{equation}
as before (see (4.25)), and the four-dimensional Einstein equation is still given by $\mathcal R_{\mu\nu} - \frac{1}{2} g_{\mu\nu} \mathcal R + \lambda g_{\mu\nu} =0$.

Now substitute (8.19) (with $L_F$ given by (8.15)) into (8.18) and contract the indices $m$ and $n$. We obtain
\begin{equation}
\mathcal R_6 - L_F + \frac{1}{2} \big( V- \frac{\partial V}{\partial h^{mn}} h^{mn} \big) + \frac{3}{2} \beta e^{{\hat\Phi}-2B} =0 \,\,,
\end{equation}
which, upon integration, gives the generalization of (4.30),
\begin{equation}
{\hat I}_{\rm bulk} = - \frac{3 \beta}{4 \kappa^2} - \frac{1}{4\kappa_{10}^2 g_s^2} \int d^{6} y \sqrt{h_6} \,\big[ V- \frac{\partial V}{\partial h^{mn}} h^{mn} \big] \,\,.
\end{equation}
Finally, substituting (8.25) into (8.23) (and using $\beta= 4 \lambda$) gives
\begin{equation}
\lambda = \frac{\kappa^2}{2} {\hat I}_{\rm brane} - \frac{\kappa^2}{8\kappa_{10}^2 g_s^2} \int d^{6} y \sqrt{h_6} \,\big[ V- \frac{\partial V}{\partial h^{mn}} h^{mn} \big] \,\,,
\end{equation}
which is just the generalization of (4.31).

\vskip 0.1cm
\hspace{-0.6cm}{\bf (A) ISD solutions}
\vskip 0.0cm

In the usual type IIB flux compactifications there is a constraint imposed on the field strength $G_3$. The Bianchi identity for ${\tilde F}_5$ combined with the noncompact components of the  Einstein equation requires \cite{28} that $G_3$ be imaginary self-dual (ISD) for a compact ${\mathcal M}_6$.\footnote{In addition to this the Bianchi identity for ${\tilde F}_5$ also requires the tadpole cancelation. This requirement can be met by introducing negative tension objects, which in our case will be either $O3$ planes of the CY orientifolds, or $D7$-branes wrapped on a four-cycle in the type IIB version of the F-theory compactifications.} Such a $G_3$ is characterized by the equations:
\begin{equation}
0=D_{\phi} W = \frac{1}{\bar \phi - \phi} \int_{{\mathcal M}_6} {\bar G}_3 \wedge \Omega \,\,,~~~~~0=D_{\tau_i} W = \int_{{\mathcal M}_6} G_3 \wedge \chi_i \,\,,
\end{equation}
where $\chi_i$ is the basis of $(2,1)$ forms on ${\mathcal M}_6$. Since the two equations in (8.27) kill resp. the $(3,0)$ and $(1,2)$ components of $G_3$, the primitive $G_3$ satisfying (8.27) contains only $(2,1)$ and $(0,3)$ components. Since this $G_3$ is ISD it satisfies $V=0$ (recall that $V \propto G_3^{\rm ISAD} \cdot {\bar G}_3^{\rm IASD}$), or equivalently ${\mathcal V}_{\rm scalar} =0$.

The compactifications using the above $G_3$ contain both supersymmetric and nonsupersymmetric solutions. To extract the supersymmetric part we need to impose further
\begin{equation}
W=D_{\rho} W=0\,\,,
\end{equation}
which kills the $(0,3)$ component of $G_3$ by (8.4), leaving only $(2,1)$ component. Thus the supersymmetry requires that $G_3$ be primitive $(2,1)$, and for this $G_3$ the scalar potential ${\mathcal V}_{\rm scalar}$ becomes necessarily of the no-scale form. Indeed, in the absence of nonperturbative corrections whole these solutions with no-scale structure are all ISD at the tree level regardless of whether or not they are supersymmetric, and for these solutions (8.26) reduces to
\begin{equation}
\lambda = \frac{\kappa^2}{2}{\hat I}_{\rm brane}\,\,.
\end{equation}
So, under (8.35), the cosmological constant problem for the ISD solutions is basically identical with that of the conventional compactifications (see (4.31)).

\vskip 0.1cm
\hspace{-0.6cm}{\bf (B) General case}
\vskip 0.0cm

We have just seen that $\lambda$ is simply given by ${\hat I}_{\rm brane}$ for the ISD (no-scale type tree level) solutions. But when the corrections enter, $V$ does not vanish anymore and $\lambda$ acquires an additional term from (8.26). The non-ISD solutions can arise by both perturbative and nonperturbative reasons. For instance one can stabilize all the moduli (including K$\ddot{\rm a}$hler) supersymmetrically by including nonperturbative contributions to the scalar potential (see \cite{29}). In this case the potential of the flux vacuum would be required not to be of no-scale type. Besides this, $V$ and consequently ${\mathcal V}_{\rm scalar}$ can fail to vanish even for ISD solutions. Though a solution is ISD at the tree level, it receives both $\alpha^{\prime}$- and $g_s$-loop corrections perturbatively. Due to these corrections, $G_3$ can acquires (1,2) and (3,0) components and as a result the solution becomes non-ISD in these cases. That is, $V$ now includes nonzero IASD contributions coming from the corrections.

For all these non-ISD solutions $\lambda$ is given by (8.26) and where the second term (the terms in the integration) is now nonzero due to the corrections $\delta V$. (8.26) is simplified if we use
\begin{equation}
\frac{\partial V}{\partial h^{mn}} h^{mn} = nV \,\,,
\end{equation}
where $n=3$ because $G_3^{\rm IASD}$ is a three-form. Indeed the K$\ddot{\rm a}$hler potential and the super potential receive the corrections of the form $\mathcal{K} = \mathcal{K}_{\rm tree} + \mathcal{K}_{p} + \mathcal{K}_{np}$ and $W= W_{\rm tree} + W_{np}$, where we see that the perturbative corrections $\mathcal{K}_{p}$ contain both $\alpha^{\prime}$- and $g_s$-loop corrections, while the superpotential receives only the nonperturbative corrections. But in any case $\mathcal{V}_{\rm scalar}$ takes the same form as (8.12) even after these corrections,\footnote{See \cite{30} for this matter. Also see \cite{30-1} for the corrections due to D3/$\,\overline{\rm D3}$-brane.} and $V$ basically satisfies (8.30) when ${\mathcal V}_{\rm scalar}$ is of no-scale type\footnote{An interesting example of this type can be found, for instance, in \cite{31}, where the factor $1/({\rm Im}\, \rho)^3$ is obtained from $e^{\mathcal{K}}$ of the no-scale potential (8.10) even without introducing the $\overline{D3}$-brane of KKLT.} and even when not.\footnote{See \cite{30} and in particular \cite{32} where it was shown that ${\mathcal V}_{\rm scalar}$ is proportional to $|W_{0}|^2$ at the minimum even under nonperturbative corrections, where $W_0$ represents the flux-induced superpotential (8.4).} By (8.17) and (8.30), (8.26) finally becomes
\begin{equation}
\lambda = \frac{\kappa^2}{2} \Big( {\hat I}_{\rm brane} + {\mathcal V}_{\rm scalar} \Big)\,\,.
\end{equation}
Now $\lambda$ consists of two parts, ${\hat I}_{\rm brane}$ and ${\mathcal V}_{\rm scalar}$, in contrast to (8.29).

\vskip 0.5cm
\hspace{-0.65cm}{\bf \large 8.3 Vanishing $\bf \lambda$}
\vskip 0.2cm

In Sec. V we have seen that $\lambda$ is forced to vanish by field equations in the conventional compactifications. This is also the case even for the flux compactifications. By (8.30), (8.24) can be rewritten as
\begin{equation}
L_F = {\mathcal R}_6 -V + \frac{3}{2} \beta e^{{\hat \Phi}-2B} \,\,.
\end{equation}
Substituting (8.32) into (8.19) then gives
\begin{equation}
T_{mn} = 2 \big( {\mathcal R}_{mn} - \frac{1}{2} h_{mn} {\mathcal R}_6 \big) + \big( V h_{mn} - 2 \frac{\partial V}{\partial h^{mn}} \big) -  \frac{3}{2} \beta e^{{\hat \Phi}-2B} h_{mn}\,\,.
\end{equation}
Again, substituting (8.33) into (8.18) gives
\begin{equation}
\big( V h_{mn} - 2 \frac{\partial V}{\partial h^{mn}} \big) -  \frac{\beta}{2}  e^{{\hat \Phi}-2B} h_{mn}=0\,\,.
\end{equation}
Finally, contracting $m$ and $n$ in (8.34) and using (8.30) gives $\beta=0$, or equivalently
\begin{equation}
\lambda=0 \,\,.
\end{equation}
Namely, $\lambda$ must also vanish in the flux compactifications as well as in the conventional compactifications.

\vskip 0.5cm
\hspace{-0.65cm}{\bf \large 8.4 Nonsupersymmetric solutions with $\bf \lambda=0$}
\vskip 0.2cm

As mentioned in Sec. 8.2 ${\mathcal V}_{\rm scalar}$ can fail to vanish even for ISD solutions. At the tree level, the $D_{\rho} W$ term in (8.12) precisely cancels the $-3 |W|^2$ term, and ${\mathcal V}_{\rm scalar}$ reduces to the no-scale form (8.10). The no-scale structure is preserved at the classical level (to the leading order in $\alpha^{\prime}$) even for nonsupersymmetric solutions for which $D_{\rho} W \propto W \neq 0$. But at the quantum level there is no guarantee that it survives perturbative (and nonperturbative) corrections. For the supersymmetric solutions such corrections can act as an F-term, and also we can have a D-term which has been ignored so far.  All these terms make nontrivial contributions to ${\mathcal V}_{\rm scalar}$, and as a result ${\mathcal V}_{\rm scalar}$ takes generically nonzero values at the quantum level. So if the contributions to $\lambda$ were coming solely from ${\mathcal V}_{\rm scalar}$, $\lambda$ would necessarily fail to vanish due to these contributions, which is precisely what happens in the ordinary flux compactifications.

In (8.31), however, $\lambda$ contains an additional term, ${\hat I}_{\rm brane}$, which is given (in the one-loop order approximation) by (see (7.2))
\begin{equation}
{\hat I}_{\rm brane}^{(1)} = \delta_0 \int r^5 dr \epsilon_5 \, \rho_T^{(1)} \,\,,
\end{equation}
where $\rho_T^{(1)}$ is arbitrary because it contains six arbitrary gauge parameters $f_m^{(0)}$ (see (6.34)). Now we decompose $\rho_T^{(1)}$ into ${\tilde \rho}_T^{(1)} + \delta \rho_T^{(1)}$ to get ${\hat I}_{\rm brane}^{(1)} \rightarrow {\tilde{\hat I}}_{\rm brane}^{(1)} + \delta {\hat I}_{\rm brane}^{(1)}$, where ${\tilde{\hat I}}_{\rm brane}^{(1)}$ and $\delta {\hat I}_{\rm brane}^{(1)}$ are given by
\begin{equation}
{\tilde{\hat I}}_{\rm brane}^{(1)}= \delta_0 \int r^5 dr \epsilon_5 \, {\tilde \rho}_T^{(1)} \,\,, ~~~~~\delta {\hat I}_{\rm brane}^{(1)} = \delta_0 \int r^5 dr \epsilon_5 \, \delta \rho_T^{(1)}\,\,.
\end{equation}
Since $\delta \rho_T^{(1)}$ is arbitrary it can be adjusted so that $\delta {\hat I}_{\rm brane}^{(1)}$ cancels (the first order deviation of) ${\mathcal V}_{\rm scalar}$. The cancelation between $\delta {\hat I}_{\rm brane}^{(1)}$ and ${\mathcal V}_{\rm scalar}$ is automatic because $\lambda$ must vanish by (8.35), and consequently (8.31) reduces to
\begin{equation}
\lambda = \frac{\kappa^2}{2} {\tilde {\hat I}}_{\rm brane} \,\,.
\end{equation}
(8.38) is the generalized version of (7.3) and where ${\tilde{\hat I}}_{\rm brane}^{(1)}$ corresponds to ${\hat I}_{\rm brane}^{(1)}$ of the conventional compactifications. ${\tilde \rho}_T^{(1)}$ in ${\tilde{\hat I}}_{\rm brane}^{(1)}$ is basically arbitrary only except for the requirement that it satisfy (7.4) to make ${\tilde{\hat I}}_{\rm brane}^{(1)}$ vanish. It can be used to break the supersymmetry at anytime we want.

Apart from this, one can also consider the case where the nonzero ${\mathcal V}_{\rm scalar}$ is compensated by the whole ${\hat I}_{\rm brane}^{(1)}$. Suppose that we have a nonsupersymmetric solution where ${\mathcal V}_{\rm scalar}$ is fine-tuned to take a nearly vanishing (or zero) value at the stable minimum. But as mentioned above ${\mathcal V}_{\rm scalar}$ does not generally survive the perturbative (both $g_s$- and $\alpha^{\prime}$-) corrections, and it necessarily acquires a nonzero value at the quantum level. In the ordinary flux compactifications this immediately leads to a nonvanishing $\lambda$ because in that case $\lambda$ is simply given by ${\mathcal V}_{\rm scalar}$. But in our case $\rho_T^{(1)}$ can be so adjusted that the nonzero ${\mathcal V}_{\rm scalar}$ is exactly compensated by the whole ${\hat I}_{\rm brane}^{(1)}$. This adjustment is automatic by (8.35) and the fact that ${\hat I}_{\rm brane}^{(1)}$ possesses gauge arbitrariness. So in this way we can obtain a nonsupersymmetric theory where $\lambda$ always vanishes even at the quantum level. In this case we do not even need ${\tilde{\hat I}}_{\rm brane}^{(1)}$ to break the supersymmetry. We can simply put ${\tilde{\hat I}}_{\rm brane}^{(1)}=0$.

So far in this section we have considered the flux compactifications of the type IIB theory to generalize the discussion on the cosmological constant problem of the conventional compactifications with $H_3 =0$. After the generalization we find that $\lambda$ appears as a sum of two terms, ${\hat I}_{\rm brane}$ and ${\mathcal V}_{\rm scalar}$, in contrasts to the case of the ordinary type IIB flux compactifications where $\lambda$ is simply given by ${\mathcal V}_{\rm scalar}$. ${\mathcal V}_{\rm scalar}$ usually receives nontrivial contributions both from perturbative and nonperturbative effects. So, even though $\lambda$ is fine-tuned to zero at the tree level, it cannot be maintained when we go up to quantum level because ${\mathcal V}_{\rm scalar}$ deviates from zero due to these corrections. Such a difficulty disappears now. In (8.31), any nonzero ${\mathcal V}_{\rm scalar}$ is always canceled by $\delta {\hat I}_{\rm brane}$ (or by the whole ${\hat I}_{\rm brane}$), and the cancelation is automatic. ${\mathcal V}_{\rm scalar}$ is just gauged away by (8.35).

\vskip 1cm
\hspace{-0.65cm}{\bf \Large {IX. Summary and discussion}}
\vskip 0.5cm
\setcounter{equation}{0}
\renewcommand{\theequation}{9.\arabic{equation}}

As an opening of the discussion on the cosmological constant problem, in the first part of this paper we studied solitonic properties of the Calabi-Yau vacua of the string theory. We first observed that the conifold singularities of the Calabi-Yau threefold can be regarded as NS-NS solitons with their masses proportional to $1/g_s^2$ because each conifold of the Calabi-Yau threefold consists of two intersecting KK-monopoles which themselves are NS-NS solitons. We then observed that the generic compact Calabi-Yau threefolds can be thought of as NS-NS objects because they usually contain certain numbers of conifolds at the singularities in such a way that the entire topology is characterized by $h^{1,1}$ and some negative $\chi$.

Such an observation coincides with the conjecture suggested in \cite{4} and \cite{5}, and may have  an important consequence in addressing the cosmological constant problem in the respect that the effect of the vacuum fluctuations exerting on the internal geometry is highly suppressed by the factor $g_s^2$ by the solitonic property of the internal dimensions. The solitonic interpretation of the internal manifolds may be extended to the whole background vacua of the various string theories. For instance in F-theory the geometry of two-dimensional transverse space of D7-brane may also be taken as an NS-NS type soliton in some sense because it does not contain the factor $g_s$ in the metric. This is interesting because the extended source of this NS-NS type soliton is D7-brane instead of (exotic) NS7-brane \cite{33}. This suggests that the D7-branes of F-theory are as rigid as the NS-NS type branes.

In the second part, we considered a configuration of a BPS D3-brane located at the conifold singularity of the Calabi-Yau threefolds to propose a new type of mechanism to address the cosmological constant problem. In Sec. IV-VII, we first considered the conventional compactifications where the $n$-form fluxes including $H_3$ are all turned off. In this case the four-dimensional cosmological constant $\lambda$ appears as a brane action density ${\hat I}_{\rm brane}$ which is basically given by a sum of two types, NS-NS type and R-R type, of vacuum energies of the brane region, and these two types of vacuum energies are forced to cancel by field equations so that $\lambda$ vanishes as a result. For the BPS state the cancelation is automatic by the supersymmetry and by field equations. But in more general cases there is an additional term in ${\hat I}_{\rm brane}$, which does not cancel by the field equations. This term, which is denoted by $\rho_T^{(1)}$, also appears in the equation of motion for $\hat{\Phi}$ and acts as a supersymmetry breaking term of the $d=4$ reduced theory.

Since the supersymmetry breaking term makes an extra contribution to $\lambda$, it (upon integration) must vanish anyhow to maintain $\lambda =0$. The field equation representing supersymmetry breaking appears in the form of the Poisson's equation for $\hat{\Phi}$. In the analogy with the ordinary electrostatics the supersymmetry breaking term $\rho_T^{(1)}$ plays the role of the charge density while $\hat{\Phi}$ plays the role of the electrostatic potential of the system. Aside from this it turns out that $\lambda$ is proportional to the total charge $Q_{\rm total}^{T}$, which is confined to the brane region and hence defined by the volume integral of $\rho_T^{(1)}$ over the brane region. So the condition $\lambda=0$ becomes equivalent to the condition $Q_{\rm total}^{T} =0$, which then implies that the $d=4$ supersymmetry remains unbroken in the bulk region (i.e., outside the brane region) because $\hat{\Phi}$ becomes a constant (zero) there by the Gauss's law of the electrostatics.

There may be many ways to satisfy $Q_{\rm total}^{T} =0$. In Sec. 7.1 we have considered two different cases in which $Q_{\rm total}^{T} =0$ is achieved in a natural way. In both cases the condition $Q_{\rm total}^{T} =0$ can be satisfied most natually by writing $\hat{\Phi}$ in terms of a Fourier series. In the case I $\hat{\Phi}$ is assumed to be a function of $r$ alone. This case includes an interesting solution where the $d=4$ supersymmetry is broken to an infinitely large extent at $r=0$. In the case II $\hat{\Phi}$ is assumed to be a function of the isometry coordinates. In our discussion we had allowed for nonzero thickness $r_B$ to the D3-brane. But in the case II we can take the limit $r_B \rightarrow 0$ whenever we want to see the thin brane features, which suggests that the whole discussion of this paper need not be restricted only to the case of the brane with nonzero thickness.

The substance of the supersymmetry breaking term is a vacuum energy density arising from the quantum excitations with components along the transverse directions. The quantum excitations induces a gauge symmetry breaking of the R-R four-form and the supersymmetry breaking occurs as a result of this gauge symmetry breaking. Since it occurs in the brane region, the brane region is locally anomalous. But the total anomaly of the brane region vanishes by the condition $\lambda =0$. The reason is because the anomaly locally occurs in the region where $\rho_T^{(1)}$ takes nonzero values. So the magnitude of the anomaly at some point of the brane region is proportional to the value of $\rho_T^{(1)}$ of that point. But $\lambda =0$ requires that $Q_{\rm total}^T$, the volume integral of $\rho_T^{(1)}$ over the whole brane region, must vanish. Thus the anomalies at each point (area) of the brane region add up to zero upon integration and the theory becomes anomaly free.

The bulk region does not suffer from this kind of anomaly because the gauge symmetry breaking of $A_4$ does not occur in the bulk region. In general the ten-dimensional superstring theory is known to be anomaly free. For instance in the type IIB theory the gravitational anomaly for the five-form ($I_A$) is canceled by the anomalies for the two left-handed Majorana-Weyl gravitinos ($2I_{3/2}$) and two right-handed Majorana-Weyl dilatinos ($-2 I_{1/2}$): $I_A + 2I_{3/2} - 2I_{1/2} =0$ \cite{21}. The anomaly of the brane region is not the usual gravitational anomaly for the five-form. Rather, it may have to be understood as a composition of the various anomalies for the string fields on the D3-brane which couple to $A_4$.

The ten-dimensional metric for the D3-brane with unbroken supersymmetry is typically given by
\begin{equation}
ds_{10}^2 = e^{-B} ds_6^2 + e^B g_{\mu\nu} dx^{\mu}dx^{\nu} \,\,,
\end{equation}
where $ds_6^2$ represents the six-dimensional internal geometry. In Sec. 7.1 it was suggested that the nonsupersymmetric theory with vanishing $\lambda$ can be easily obtained from (9.1) by replacing $ds_6^2$ by $e^{\hat \Phi} ds_6^2$ where $\hat \Phi$ is a solution to the poisson's equation (6.33). In the bulk region the supersymmetry is preserved (${\hat \Phi} =0$) so the metric $ds_{10}^2$ is still given by (9.1). But in the brane region the supersymmetry is broken (${\hat \Phi} \simeq g_s U_{(1)}$) and therefore $ds_{10}^2$ will take the form
\begin{equation}
ds_{10}^2 \cong e^{-B} (1+ g_s U_{(1)}) ds_6^2 + e^B g_{\mu\nu} dx^{\mu}dx^{\nu} \,\,.
\end{equation}
The additional term $g_s U_{(1)}$ in (9.2) is of course due to the supersymmetry breaking.

Still in the conventional compactifications the supersymmetry breaking gives a mass to the dilaton because the supersymmetry breaking term generates a potential for the dilaton. From (4.1) and (6.8) one finds the action for the dilaton (in the brane region) is given by
\begin{equation}
I(\Phi) = \frac{1}{2 \kappa_{10}^2 g_s^2} \int d^4 x \sqrt{-g_4} \int d^6 y \sqrt{h_6}
\Big[ 4 \, e^{\hat{\Phi}-2B} (\partial \Phi)_x^2 + V(\Phi) + \cdots \, \Big] \,\,,
\end{equation}
where $(\partial \Phi)_x^2 \equiv g^{\mu\nu} \partial_{\mu} \Phi \partial_{\nu} \Phi$, and (from the leading term of $\delta \mu_{\rm T}^m (\Phi)$ in (6.29)) the potential $V(\Phi)$ takes the form
\begin{equation}
V(\Phi) = 2 \kappa_{10}^2 \, \delta_0 \, \rho_T^{(1)} e^{\Phi} + \cdots \,\,.
\end{equation}
From $\partial^2 V(\Phi) / \partial \Phi^2$ the characteristic scale of the dilaton mass is estimated to be $m_{\Phi}^2 \approx c_0 \rho_T^{(1)}$, where $c_0$($\equiv$$\,2 \kappa_{10}^2 g_s \delta_0$) becomes $c_0 \simeq (2 \pi)^7 g_s l_s^8 /r_B^6$ upon setting $2 \kappa_{10}^2 = (2 \pi)^7 l_s^8$ where $l_s$ is the fundamental scale of the string theory, $l_s = 1/m_s = \sqrt{\alpha^{\prime}}$. Also the thickness (the characteristic size) of the brane should be of order $l_s$: $r_B \sim l_s$, so we have $c_0 \sim (2 \pi)^7 g_s l_s^2$ and therefore
\begin{equation}
m_{\Phi}^2 \simeq (2 \pi)^7 g_s \frac{\rho_T^{(1)}}{m_s^2} \,\,.
\end{equation}

In (9.5) the magnitude of $\rho_T^{(1)}$ is given by $\rho_T^{(1)} \sim \nu_1^m$, and where it is natural to assume that $\nu_1^m$ is of the same order as $\mu_1$, which is the first order correction to $\mu_0$ in the $g_s$ expansion. Since $\mu_0 \sim m_s^4$ (more precisely, it is $\mu_0 \sim m_s^4 / (2\pi)^3$) we also expect $\mu_1 \sim \nu_1^m \sim m_s^4$ and consequently $\rho_T^{(1)} \sim m_s^4$. Putting all these together (and omitting the factor $(2 \pi)^7$) we obtain $m_{\Phi}^2 \sim g_s m_s^2$ from (9.5). The dilaton mass $m_{\Phi}$ gives a typical mass scale for the supersymmetry breaking and it may be roughly identified with the mass scale of the Standard Model superpartners, $m_{\rm sp}$ \cite{20}. We finally have
\begin{equation}
m_{\rm sp}^2 \sim g_s m_s^2 \,\,.
\end{equation}
(9.6) suggests that the magnitude of $m_{\rm sp}$ could be much larger than the conventional LHC scale of order $\sim TeV$.

As a final discussion on the cosmological constant problem of the conventional compactifications it was argued that the configuration with an unbroken supersymmetry is as equally possible as the configuration with broken supersymmetry but the latter is more favored by the action principle than the former. This argument, however, is valid only in the realistic models in which the field equations are sufficiently relaxed, and in addition to this there is some subtlety in writing down the classical action for the theory containing self-dual five-form as a field content. For this matter of supersymmetry breaking there is a different viewpoint that $f_m^{(0)}$ must be taken as a nonzero function from the beginning. In this viewpoint $f_m^{(0)}$ is not just an arbitrary gauge parameter. Rather, it is a distribution function describing charge (vacuum energy) configurations inside the brane region. So $\rho_T^{(1)}$ does not vanish because $\nu_1^m$ are nonzero constants, and therefore the $d=4$ supersymmetry is always broken and the solutions with an unbroken supersymmetry do not exist from the beginning.

These are the whole story of the cosmological constant problem of the conventional compactifications. In Sec. VIII, which is the last section of the second part, the above discussions have been generalized to the case of the flux compactifications of type IIB theory where the $n$-form fluxes are all turned on to stabilize the moduli. In this generalized theory we found that $\lambda$ appears as a sum of two terms, ${\hat I}_{\rm brane}$ and ${\mathcal V}_{\rm scalar}$, which contrasts with the ordinary type IIB flux compactifications where $\lambda$ is simply given by ${\mathcal V}_{\rm scalar}$. In the ordinary flux compactifications ${\mathcal V}_{\rm scalar}$ can be fine-tuned to zero at the classical level to obtain a theory with vanishing $\lambda$. But ${\mathcal V}_{\rm scalar}$ necessarily deviates from zero because it receives non-trivial contributions coming from loops of bulk fields or from any other kind of perturbative or nonperturbative effect that one would need to include in order to stabilize the moduli of the Calabi-Yau threefold. As a result, $\lambda =0$ cannot be maintained at the quantum level in the ordinary flux compactifications.

Indeed, the vanishing $\lambda$ for the nonsupersymmetric vacua usually depends on the tree level structure of the K$\ddot{\rm a}$hler potential for the K$\ddot{\rm a}$hler moduli. But since this K$\ddot{\rm a}$hler potential is unstable against both perturbative and nonperturbative corrections, so is the tree level potential ${\mathcal V}_{\rm scalar}$. Thus even though we fine-tune ${\mathcal V}_{\rm scalar}$ so that it vanishes at the stable minimum, it cannot be maintained once the corrections enter because ${\mathcal V}_{\rm scalar}$ necessarily acquires nonzero values at the stable minimum due to corrections. So in the ordinary flux compactifications $\lambda =0$ cannot be maintained because $\lambda$ is simply given by ${\mathcal V}_{\rm scalar}$ there.

Such a thing can be avoided in our case. First, in our case $\lambda$ appears as a sum of two terms, ${\hat I}_{\rm brane}$ and ${\mathcal V}_{\rm scalar}$, not just ${\mathcal V}_{\rm scalar}$ alone. Further, ${\hat I}_{\rm brane}$ possesses gauge arbitrariness as mentioned in Sec. 8.4. Finally $\lambda$ is forced by (8.35) to vanish. So we can always maintain $\lambda =0$ regardless of whether ${\mathcal V}_{\rm scalar}$ acquires nonzero values at the stable minimum or not because any nonzero values of ${\mathcal V}_{\rm scalar}$ is automatically gauged away (namely cancel with ${\hat I}_{\rm brane}$) by (8.35). So one of the simple way to obtain a nonsupersymmetric theory with $\lambda =0$ is just to find a tree level solution where $D_{\rho} W \propto W \neq 0$ and ${\mathcal V}_{\rm scalar}$ is fine-tuned to take a nearly vanishing value at the stable minimum. Then the nonzero values of ${\mathcal V}_{\rm scalar}$ arising from the higher order (both of $\alpha^{\prime}$- and $g_s$-) corrections (and the small corrections coming from nonperturbative effects) are all gauged away and we readily obtain a theory with broken supersymmetry where $\lambda$ always vanishes even at the quantum level.

\vskip 1cm
\begin{center}
{\large \bf Acknowledgement}
\end{center}

This work was supported by the National Research Foundation of Korea (NRF), under Grant No. 353-2009-2-C00045, funded by the Korean Government.


\vskip 1cm

\begin{thebibliography}{999}

\bibitem{1} P. Candelas, P. S. Green and T. H$\ddot{\rm u}$bsch, {\it Rolling among Calabi-Yau vacua}, Nucl. Phys. {\bf B330} (1990) 49-102.

\bibitem{1-1} P. S. Green and T. H$\ddot{\rm u}$bsch, {\it Possible phase transitions among Calabi-Yau compactifications}, Phys. Rev. Lett. {\bf 61} (1988) 1163-1166; {\it Connecting moduli spaces of Calabi-Yau threefolds}, Comm. Math. Phys. 119 (1989) 431-441;\\ P. Candelas, P. S. Green and T. H$\ddot{\rm u}$bsch, {\it Finite distance between distinct Calabi-Yau manifolds}, Phys. Rev. Lett. {\bf 62} (1989) 1956-1959.

\bibitem{2} M. R. Douglas and G. Moore, {\it  D-branes, Quivers, and ALE instantons}, [hep-th/9603167];\\ A. Kehagias, {\it  New type IIB vacua and their F-Theory interpretation}, Phys. Lett. {\bf B435} (1998) 337-342 [hep-th/9805131];\\ I. R. Klebanov and E. Witten, {\it Superconformal field theory on threebranes at a Calabi-Yau singularity}, Nucl. Phys. {\bf B536} (1998) 199-218 [hep-th/9807080];\\ D. R. Morrison and M. R. Plesser, {\it Non-spherical horizons, I}, Adv. Theor. Math. Phys. {\bf 3} (1999) 1-81 [hep-th/9810201];\\ I. R. Klebanov and A. A. Tseytlin, {\it Gravity duals of supersymmetric $SU(N) \times SU(N+M)$ gauge theories}, Nucl. Phys. {\bf B578} (2000) 123-138 [hep-th/0002159];\\  L. A. Pando Zayas and A. A. Tseytlin, {\it 3-branes on resolved conifold}, J. High Energy Phys. {\bf 11} (2000) 028 [hep-th/0010088].

\bibitem{3} K. Dasgupta and S. Mukhi, {\it Brane constructions, conifolds and M-Theory}, Nucl. Phys. {\bf B551} (1999) 204-228 [hep-th/9811139].

\bibitem{4} E. K. Park and P. S. Kwon, {\it A comment on $p$-branes of ($p+3$)$\rm{d}$ string theory}, J. High Energy Phys. {\bf 05} (2009) 057 [arXiv:0812.0227].

\bibitem{5} E. K. Park and P. S. Kwon, {\it NS-branes in 5d brane world models}, Phys. Rev. {\bf D82} (2010) 046001 [arXiv:1007.1290].

\bibitem{6} M. Bershadsky, V. Sadov and C. Vafa, {\it D-strings on D-manifolds}, Nucl. Phys. {\bf B463} (1996) 398-414 [hep-th/9510225];\\ Also see, M. Aganagic, A. Karch, D. Lust and A. Miemiec, {\it Mirror symmetries for brane configurations and branes at singularities}, Nucl. Phys. {\bf B569} (2000) 277-302 [hep-th/9903093];\\J. McOrist and A. B. Royston, {\it Relating conifold geometries to NS5-branes}, Nucl. Phys. {\bf B849} 2011 573-609 [arXiv:1101.3552]; {\it T-dualising the deformed and resolved conifold}, [arXiv:1107.5895].

\bibitem{7} E. Bergshoeff, R. Kallosh and T. Ortin, {\it Duality versus supersymmetry and compactification}, Phys. Rev. {\bf D51} (1995) 3009-3016 [hep-th/9410230].

\bibitem{8} K. Ohta and T. Yokono, {\it Deformation of conifold and intersecting branes}, J. High Energy Phys. {\bf 02} (2000) 023 [hep-th/9912266].

\bibitem{9} See, for instance, C. V. Johnson, {\it D-branes}, Cambridge University Press (2003).

\bibitem{10} D. J. Gross and M. J. Perry, {\it Magnetic monopoles in Kaluza-Klein theories}, Nucl. Phys. {\bf B226} (1983) 29-48.

\bibitem{11} T. Eguchi and A. J. Hanson, {\it Asymptotically flat self-dual solutions to euclidean gravity}, Phys. Lett. {\bf B74} (1978) 249-251.

\bibitem{12} M. K. Prasad, {\it Equivalence of Eguchi-Hanson metric to two center Gibbons-Hawking metric}, Phys. Lett. {\bf B83} (1979) 310.

\bibitem{13} For the review of this matter, see for instance. S. Weinberg, {\it The cosmological constant problem}, Rev. Mod. Phys. {\bf 61} (1989) 1-21.

\bibitem{14} V. A. Rubakov and M. E. Shaposhnikov, {\it Extra space-time dimensions: towards a solution of the cosmological constant problem}, Phys. Lett. {\bf B125} (1983) 139-143.

\bibitem{15} C. Schmidhuber, {\it Micrometer gravitinos and the cosmological constant}, Nucl. Phys. {\bf B585} (2000) 385-394 [hep-th/0005248].

\bibitem{16} See, for instance, E. K. Park and P. S. Kwon, {\it  A self-tuning mechanism in (3+p)d gravity-scalar theory}, J. High Energy Phys. {\bf 11} (2007) 051 [hep-th/0702171].

\bibitem{17} M. J. Duff and J. X. Lu, {\it The self-dual type IIB superthreebrane}, Phys. Lett. {\bf B273} (1991) 409-414.

\bibitem{18} M. J. Duff, R. R. Khuri and J. X. Lu, {\it String solitons}, Phys. Rept. {\bf 259} (1995) 213-326 [hep-th/9412184].

\bibitem{19} S. Deser and R. Jackiw, {\it Three-dimensional cosmological gravity: Dynamics of constant curvature}, Annals Phys., {\bf 153} (1984) 405-416.

\bibitem{20} J. Polchinski, {\it String theory}, Vols. I and II; Cambridge University Press (1998);\\ Also see, J. Polchinski, {\it Dirichlet-Branes and Ramond-Ramond Charges}, Phys. Rev. Lett. {\bf 75} (1995) 4724-4727 [hep-th/9510017].

\bibitem{21} See, for instance, K. Becker, M. Becker and J. H. Schwarz, {\it String theory and M-Theory}, Cambridge University Press (2007).

\bibitem{22} L. Alvarez-Gaume and E. Witten, {\it Gravitational anomalies}, Nucl. Phys. {\bf B234} (1984) 269-330.

\bibitem{23} C. Schmidhuber, {\it Brane supersymmetry breaking and the cosmological constant: Open problems}, Nucl. Phys. {bf B619} (2001) 603-614 [hep-th/0104131].

\bibitem{24} S. Kachru, J. Kumar and E. Silverstein, {\it Vacuum energy cancellation in a non-supersymmetric string}, Phys. Rev. {\bf D59} (1999) 106004 [hep-th/9807076].

\bibitem{25} S. Kachru and E. Silverstein, {\it Self-dual nonsupersymmetric type II string compactifications}, J. High Energy Phys. {\bf 11} (1998) 001 [hep-th/9808056];\\ J. A. Harvey, {\it String duality and non-supersymmetric strings}, Phys. Rev. {\bf D59} (1999) 026002 [hep-th/9807213];\\
    G. Shiu and S.-H. Henry Tye, {\it Bose-Fermi Degeneracy and Duality in Non-Supersymmetric Strings}, Nucl. Phys. {\bf B542} (1999) 45-72 [hep-th/9808095];\\  S. Kachru and E. Silverstein, {\it On Vanishing Two Loop Cosmological Constants in Nonsupersymmetric Strings}, J. High Energy Phys. {\bf 01} (1999) 004 [hep-th/9810129];\\ R. Blumenhagen and L. G$\ddot{\rm o}$erlich, {\it Orientifolds of Non-Supersymmetric Asymmetric Orbifolds}, Nucl. Phys. {bf B551} (1999) 601-616 [hep-th/9812158];\\ C. Angelantonj, I. Antoniadis and K. Foerger, {\it Non-Supersymmetric Type I Strings with Zero Vacuum Energy}, Nucl. Phys. {\bf B555} (1999) 116-134 [hep-th/9904092];\\ I. Antoniadis, E. Dudas and A. Sagnotti, {\it Brane Supersymmetry Breaking}, Phys. Lett. {\bf B464} (1999) 38-45 [hep-th/9908023].

\bibitem{26} S. Gukov, C. Vafa and E. Witten, {\it CFT's From Calabi-Yau Four-folds}, Nucl. Phys. {\bf B584} (2000) 69-108 [hep-th/9906070].

\bibitem{27} S. Kachru, M. Schulz and S. P. Trivedi, {\it Moduli Stabilization from Fluxes in a Simple IIB Orientifold}, J. High Energy Phys. {\bf 10} (2003) 007 [hep-th/0201028].

\bibitem{28} S. B. Giddings, S. Kachru and J. Polchinski, {\it Hierarchies from Fluxes in String Compactifications}, Phys. Rev. {\bf D66} (2002) 106006 [hep-th/0105097].

\bibitem{29} E. Witten, {\it Non-Perturbative Superpotentials In String Theory}, Nucl. Phys. {\bf B474} (1996) 343-360 [hep-th/9604030];\\ S. Kachru, R. Kallosh, A. Linde and S. P. Trivedi, {\it de Sitter Vacua in String Theory}, Phys. Rev. {\bf D68} (2003) 046005 [hep-th/0301240];\\ See also M. R. Douglas and S. Kachru, {\it Flux Compactification}, Rev. Mod. Phys. {\bf 79} (2007) 733-796 [hep-th/0610102].

\bibitem{30} For the loop corrections in type IIB flux compactifications, see M. Cicoli, J. P. Conlon and F. Quevedo, {\it Systematics of String Loop Corrections in Type IIB Calabi-Yau Flux Compactifications}, J. High Energy Phys. {\bf 01} (2008) 052 [arXiv:0708.1873];\\ Also see O. DeWolfe and S. B. Giddings, {\it Scales and hierarchies in warped compactifications and brane worlds}, Phys. Rev. {\bf D67} (2003) 066008 [hep-th/0208123];\\ K. Becker, Yu-Chieh Chung, Guangyu Guo, {\it Metastable Flux Configurations and de Sitter Spaces}, Nucl. Phys. {\bf B790} (2008) 240-257 [arXiv:0706.2502].

\bibitem{30-1} O. DeWolfe, L. McAllister, G. Shiu and B. Underwood, {\it D3-brane Vacua in Stabilized Compactifications}, J. High Energy Phys. {\bf 09} (2007) 121 [hep-th/0703088].

\bibitem{31} A. Saltman, E. Silverstein, {\it The Scaling of the No Scale Potential and de Sitter Model Building}, J. High Energy Phys. {\bf 11} (2004) 066 [hep-th/0402135].

\bibitem{32} K. Becker, M. Becker, K. Dasgupta and S. Prokushkin, {\it Properties Of Heterotic Vacua From Superpotentials}, Nucl. Phys. {\bf B666} (2003) 144-174 [hep-th/0304001];\\  G. L. Cardoso, G. Curio, G. Dall'Agata and D. L$\ddot{\rm u}$st, {\it BPS Action and Superpotential for Heterotic String Compactifications with Fluxes}, J. High Energy Phys. {\bf 10} (2003) 004 [hep-th/0306088];\\ S. Gurrieri, A. Lukas and A. Micu, {\it Heterotic string compactified on half-flat manifolds}, Phys. Rev. {\bf D70} (2004) 126009 [hep-th/0408121];\\  D. L$\ddot{\rm u}$st, S. Reffert, E. Scheidegger, W. Schulgin and S. Stieberger, {\it Moduli Stabilization in Type IIB Orientifolds (II)}, Nucl. Phys. {\bf B766} (2007) 178-231 [hep-th/0609013].

\bibitem{33} E. Eyras and Y. Lozano, {\it Exotic branes and nonperturbative seven branes}, Nucl. Phys. {\bf B573} (2000) 735 [hep-th/9908094].

\end{thebibliography}
\end{document}